\documentclass[twocolumn]{aastex631}

\newcommand\decl{decl.}
\newcommand\deltadecl{$\delta \hbox{decl.}$}
\newcommand\RA{R.A.}
\newcommand\deltaRA{$\delta \hbox{RA}$}

\defcitealias{Lubow2021}{Paper 1}

\graphicspath{{./}{figures/}}

\journalinfo{Accepted for publication in the Astronomical Journal}

\received{December 7, 2021}
\revised{June 2, 2022}
\accepted{June 19, 2022}
\submitjournal{the Astronomical Journal}
\shorttitle{ Improvements to PS1 Astrometry: II. Corrections for
Differential Chromatic Refraction }
\shortauthors{White, Lubow, \& Shiao}

\begin{document}

\title{Improvements to Pan-STARRS1 Astrometry: II. Corrections for
Differential Chromatic Refraction}

\correspondingauthor{Richard L. White}
\email{rlw@stsci.edu}

\author[0000-0002-9194-2807]{Richard L. White}
\author[0000-0002-4636-7348]{Stephen H. Lubow}
\author[0000-0001-7842-3714]{Bernie Shiao}
\affiliation{Space Telescope Science Institute, 3700 San Martin Drive, Baltimore, MD 21218, USA}



\begin{abstract}
In a previous paper, we applied the Gaia DR2 catalog to improve the astrometric accuracy of about 1.7 billion objects in Pan-STARRS1 Data Release 2 (PS1 DR2).
We report here on further improvements made by utilizing Gaia EDR3 and correcting for the effects of
differential chromatic refraction (DCR) in declination.  We extend the correction algorithm in Paper 1 by iteratively subtracting color- and declination-dependent PS1/Gaia EDR3 declination residuals.
We determine the astrometric improvement for $\sim440$ million reference objects that are point-like and cross-match to Gaia EDR3.
For this set of objects, Gaia EDR3 provides a $\sim3$\% improvement in PS1 astrometry over Gaia DR2, and DCR corrections provide an additional $\sim5$\% improvement.
DCR corrections increase substantially for objects observed away from the zenith.  DCR corrections lead to
an astrometric improvement of $\sim30$\% for blue objects ($0<g-i<1$)
that are $50^\circ$ away from the zenith.
The amplitude of
systematic astrometric errors from these effects is substantially
reduced to less than 1~mas for objects with PS1 colors in the range
$0 < g-i < 4.5$, which makes this a useful
astrometric reference catalog in fields where there are few Gaia
stars.  The improved astrometric data will be available through the 
Mikulski Archive for Space Telescopes PS1 catalog interfaces.

\end{abstract}



\section{Introduction}

The Panoramic Survey Telescope and Rapid Response System (Pan-STARRS) at Haleakala Observatory in Hawaii was used to carry out the Pan-STARRS1 survey (PS1) mainly from 2010 to 2014.
PS1 covers a region north of declination ${-}30$ degrees that constitutes about 75\% of the sky.
This region was covered about 12 times
in each of five broadband filters  ($g$, $r$, $i$, $z$, $y$).
\cite{Chambers2016} provides an overview of PS1.
Data releases DR1 and DR2 have been made.
DR1 (2016 December) contained only average information resulting from individual exposures. The second data release, DR2 (2019 January), contains time-dependent information obtained from individual exposures.
In this paper we use data from DR2.\footnote{When we refer to PS1 data products, we mean the DR2 versions unless we explicitly mention
DR1.}

In a recent paper \citep[][hereafter Paper 1]{Lubow2021}, we improved the astrometric accuracy of 1.7 billion
PS1 objects  that have more than two detections by using Gaia DR2 as an astrometric reference \citep{Prusti2016,Lindegren2018}. A subset, consisting of about 440 million of these PS1 objects, are called {\it reference objects}.
They are objects that have more than two detections, are point-like, and cross-match to Gaia.
The point-like quality is determined by requiring a point-source score greater than 0.9 on a scale that ranges from 0 (extended) to 1 (point source) \citep[][ \url{https://archive.stsci.edu/prepds/ps1-psc/}]{Tachibana2018}.
To perform the cross-matching, for each Gaia source, we determine
the nearest PS1  object that satisfies the detection and point-like requirements within a 2 arcsecond search radius  without accounting for corrections due to proper motions and parallaxes. These PS1 cross-matched objects are the reference objects. In \citetalias{Lubow2021}, the cross-matching used Gaia DR2,
while this paper uses Gaia EDR3.

In \citetalias{Lubow2021}, we found that there are spatially correlated PS1/Gaia DR2 astrometric residuals of the reference objects on the $\sim 1$ arcmin scale and employed an algorithm for reducing these residuals.
For each PS1 object, the algorithm applies an astrometric correction that is the median  PS1 to Gaia shift of the nearest 33 reference objects, excluding the object being corrected.
In addition we determined the proper motions of these PS1 objects (using only measurements from PS1) and applied similar corrections to these proper motions.
The median astrometric error for the reference objects was reduced by about 33\% in position to 9.0 mas and about 24\% in proper motion to 4.8 mas/yr.

After applying these corrections,
we found that within PS1 stripes (bands of \decl\ that range in size between 3 to 10 degrees) there are systematic variations of the PS1/Gaia \decl\ residuals as a function of the Gaia  $b-r$ color. They increase
with angle away from zenith as the airmass increases (scaling as the tangent of the zenith distance) and are largest for bluer colors (see Figures 18 and 19 in \citetalias{Lubow2021}). The residuals are caused by changes in atmospheric refraction with color, an effect called
differential chromatic refraction (DCR).

DCR corrections have been previously made by \cite{Magnier2020}, who used ground-based reference objects from 2MASS.
Their corrections are included in the PS1 positions that are used as inputs to our correction algorithms.
For the \cite{Magnier2020} DCR corrections, both the 2MASS astrometric reference objects and the PS1 objects experience DCR.
The color difference between each PS1 object and its reference stars
was taken into account in making the DCR correction.  The color and declination-dependent errors described in \citetalias{Lubow2021} are the small residual errors that
remain after the corrections that were applied in constructing the PS1 catalog.

In this paper we describe improvements
to the astrometry of the PS1 DR2 catalog by using Gaia EDR3 \citep{GaiaEDR3} together with DCR corrections.
Some improvements occur through the use of the newer version of Gaia, which has more reference objects along with more accurate
positions and proper motions. However, larger improvements are made through the DCR corrections, particularly for objects observed well away from the zenith.
Since Gaia is unaffected by DCR, only the DCR effects of PS1 need to be considered. The DCR corrections we apply with Gaia are then absolute, not relative
to the color of the Gaia reference objects, which provides a significant advantage compared with the original correction using 2MASS.
The outline of this paper is as follows.
In Section \ref{sec:algorithm} we describe the correction algorithm.
Section \ref{sec:results} describes the results of applying the corrections.
Section \ref{sec:errors} presents a detailed assessment of the errors in the corrected positions.
Section  \ref{sec:summary} summarizes our results.

\section{Correction Algorithm}
\label{sec:algorithm}

As in \citetalias{Lubow2021}, we apply the astrometric corrections in a database system.
The PS1 database contains the tables that hold the information about detected objects, such as
positions and magnitudes. \cite{Flewelling2020} and the PS1 archive documentation (\url{https://panstarrs.stsci.edu}) describe the database structure in detail.
The \textit{Detection} table (actually, a view or virtual table) contains positional information based on individual single-epoch exposures.
The mean positions and epochs are found in the \textit{ObjectThin} table, and
the mean fluxes and magnitudes, both of which are
determined from the \textit{Detection} table measurements, are found in the \textit{MeanObject} table.
Stack images are produced by combining all the single-epoch exposures in a given filter to obtain a
deeper image.
Positional and magnitude
information on each object for each filter as determined from the stack images are found in the  \textit{StackObjectThin} table.

To carry out both the PS1 to Gaia shift correction of nearby reference objects described in Paper 1  and the DCR correction, we first apply the Paper 1 correction.   Using the color-based position residuals relative to Gaia, we then apply a DCR correction.  However, these two steps alone are insufficient
because the reference objects used in the first step have not been corrected for DCR. (The implications of this inconsistency are described at the end of this section.)  To remedy this situation, we apply an iteration scheme that is described in Section \ref{sec:steps}.

For objects observed close to the meridian, refraction effects are much larger in \decl\ than in \RA\
Most PS1 fields are, in fact, observed close to the meridian. In practice we find that
the DCR corrections can be modeled as shifts in \decl\ only that are a
function \decl\ and color (see Section~\ref{sec:errors} for analysis supporting this approach).
These corrections are determined by the PS1/Gaia EDR3 \decl\ residual distributions as a function of color within each stripe.
The DCR corrections are carried out in a series of nine iterations that are numbered 0 to 8.
The first iteration  (iteration 0) applies the correction algorithm of \citetalias{Lubow2021} only, which made
no color corrections at all.
After this iteration, the  \decl\ residuals are determined as a function of color. These residuals are used to
provide DCR corrections for iteration 1.
In iteration 1, the  algorithm of \citetalias{Lubow2021} is applied with PS1 positions that are DCR corrected.
The iterations continue by applying the algorithm of \citetalias{Lubow2021} with initial PS1 positions that are DCR corrected based on
the cumulative color-based \decl\ residuals from previous iterations.

\subsection{Details of the iterative algorithm}
\label{sec:steps}

In  more detail, the steps to correct a stripe are as follows:

1. To begin iteration 0, we determine the $g-i$ color of PS1 reference objects in the stripe based on mean object information in columns \textit{gMeanPSFMag} and \textit{iMeanPSFMag} in table
\textit{MeanObject}, if available. If that color information is not available, then we use stack object information from  columns \textit{gPSFMag} and \textit{iPSFMag} in table
\textit{StackObjectThin}, if that is available. If neither is available, then that reference object is removed from the list of reference objects.
Only 0.07\% of the reference objects are removed (as expected, since the Gaia reference stars are usually bright enough to be detected by PS1).

2. For each reference PS1 object, we compute the Gaia \RA\ and \decl\ at the PS1 \RA\ and \decl\ mean detection epochs, respectively (\textit{mdmjdra}, \textit{mdmjddec}), of the cross-matched Gaia source by using the Gaia proper motions and parallaxes. We determine the Gaia EDR3/PS1 \RA\ and \decl\ residuals
of the reference objects that we call the {\it initial residuals}.

3. We apply the median neighbor shift correction algorithm of \citetalias{Lubow2021} to these initial reference object residuals. After this correction
we determine the post-correction residuals in \RA\ and \decl\ between
Gaia and PS1. This step completes iteration 0 for the stripe.

4. In iteration 1, we create a DCR histogram table $H$ of the post-correction iteration 0 \decl\ reference object residuals as a function of PS1 $g-i$ color. The histogram is stored as a database
table with a row for each bin that contains the minimum and maximum color, along with the median of the \decl\ residuals, \deltadecl.
There are initially 100 bins in $g-i$ with an equal number number of objects for $g-i < 3.0$.
The bluest (smallest $g-i$) bin is then refined by a factor of 30 to have an equal number of objects in each bin.
Bins for $g-i \ge 3.0$ (very red objects) are added and contain an equal number of objects as the refined bluest bins.  These refinements are done because
of the relatively small number of objects outside the color range of bin 2 to 100.
The maximum and minimum bin color values do not change across iterations within a stripe, but vary across stripes.
For each bin (row), we store in the \deltadecl\ column the median of the iteration 0 \decl\ residuals for the color range in that bin.

5.  In iterations greater than 1, we create a new, temporary DCR histogram table $HT$ of the \decl\ reference object residuals from the previous iteration as a function of PS1 $g-i$
using the same bins as used in histogram $H$.
We increment \deltadecl\ in each bin of histogram $H$ by the value of \deltadecl\ in the same bin of the $HT$ histogram.
To speed up convergence for objects with $g-i \ge 3.0$, we multiply the residuals in $HT$ by a factor of 1.5 and add that quantity to the \deltadecl\ value in the DCR histogram $H$.

6. For each object, we determine the bin (table row) for which its $g-i$ color lies between the minimum and maximum color in the DCR histogram.
For each such object, we subtract the \deltadecl\ value for that bin from the initial \decl\ residual to produce DCR corrected initial residuals.

7. We apply the median neighbor shift correction algorithm of \citetalias{Lubow2021}.

8. After iteration 1, we repeat steps 5, 6, and 7 six more times for reference objects only, providing iterations 2 through 7.

9. In iteration 8, we apply step 5 for reference objects and then run steps 6 and 7 for all 1.7 billion PS1 objects with more than two detections.
If  PS1 $g-i$ colors are unavailable for an object, then its DCR \decl\ correction is set to zero.  About 11\% of the objects do not have PS1 $g-i$ colors.
Note that objects that do not have $g-i$ colors are typically red and so are less affected by atmospheric refraction.

10. We add the Gaia minus PS1 residuals from iteration 8 to the initial PS1 positions to produce a table of corrected PS1 positions.

\begin{figure*}
\centering
\includegraphics[width=\textwidth]{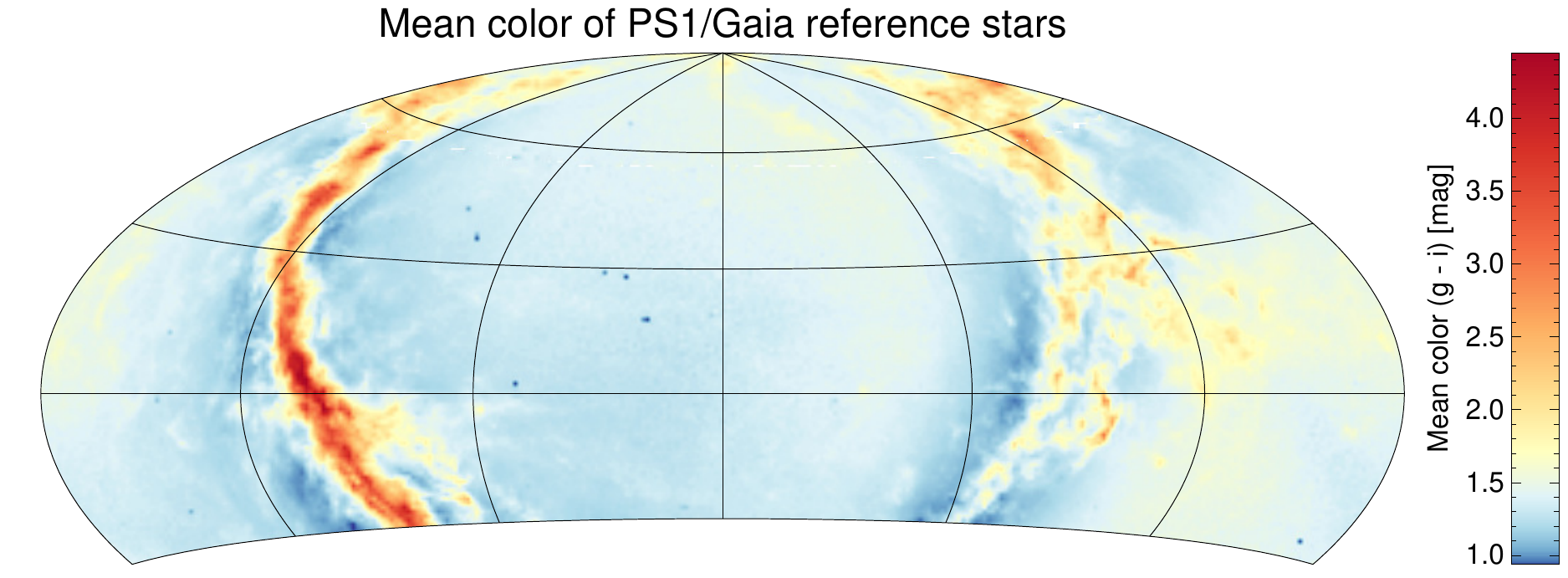}
\caption{Mean $g-i$ color of PS1/Gaia reference stars as a function
of \RA\ and decl.  The lower \decl\ limit of the PS1 survey is
$\delta = -30^\circ$. Over most of the sky the color is relatively
uniform, but dust absorption in the plane of the Galaxy leads to
extremely red mean colors.  Note also the blue halo slightly off
the plane, which is the population of younger stars near the plane
peeking out above the edge of the dust. The blue spots at higher
latitudes are globular clusters. These color differences
lead to the need for an iterative solution to the DCR correction.
The object density is much higher in regions of unusual color, so
the effect of color variations on the convergence is large despite
the small sky area covered by the plane.
}
\label{fig:skycolor}
\end{figure*}

\subsection{Convergence of the iteration}
\label{sec:convergence}

We experimented with different numbers of iterations and found that the convergence continued somewhat up to iteration 8, especially at the red end, $g-i> 4$.
Further iterations did not lead to significant improvement.
In making the DCR corrections, we assume that that the objects are observed on or close to the meridian.
Consequently, we did not apply DCR corrections to \RA, which would be largely unaffected by DCR.
Instead, for \RA, we only used the median neighbor shift correction algorithm of \citetalias{Lubow2021}.
Proper motions do not benefit from DCR corrections, assuming that the object color does not change across epochs of observations
(and that our assumption that objects are observed as they transit the meridian is correct).
We therefore only applied the median neighbor shift correction algorithm of \citetalias{Lubow2021} with Gaia EDR3 to the proper motions.

Notice that the algorithm above does not iterate on the median neighbor shift corrected positions. Instead, it starts each iteration
with DCR corrections to the initial (not median neighbor corrected) positions of the PS1 objects. The reason is that
we found that iterating on the median neighbor corrected positions can sometimes lead to instability, resulting in a lack of convergence.

The DCR histograms are constructed for reference objects within a stripe, leading to corrections that are constant across
the stripe for objects of a given color.
We experimented with interpolating the corrections as a function of \decl\ within the stripe using the corrections from neighboring stripes. However, that refinement did not
lead to better astrometric accuracy and consequently we did not adopt that approach.

We discuss here  the need for the iteration between the median neighbor shift correction of Paper 1 and  the DCR correction. The iteration is needed because reference star
and object colors vary systematically as a function of position on
the sky (see Fig.~\ref{fig:skycolor}).  That means, for example,
that many of the red objects (near the Galactic plane) also have
red reference stars.  That makes the initial positions (with no DCR
corrections) more accurate for those objects.  But it also means
that our DCR correction in the first iteration underestimates the
true amplitude of the DCR bias.  An isolated red star far from the
Galactic plane (with average color reference stars) therefore has
a larger error than is estimated from the measured mean error for
stars of that color over the sky (which is dominated by stars in
the plane).  After repeated iterations, the model for the ``true''
DCR correction gradually improves, leading to more uniform residual
errors over the sky.

\subsection{Database implementation}
\label{sec:implementation}

The above steps were carried out in a Microsoft SQL Server database using the JHU spherical library \citep{Budavari2010} for finding nearest neighbors and using Common Language Runtime (CLR) functions
for computing the median values and Gaia parallax shifts. The running time for all PS1 objects we consider was about 10 days.

\section{Results}
\label{sec:results}

\begin{figure}
\includegraphics[width=\columnwidth]{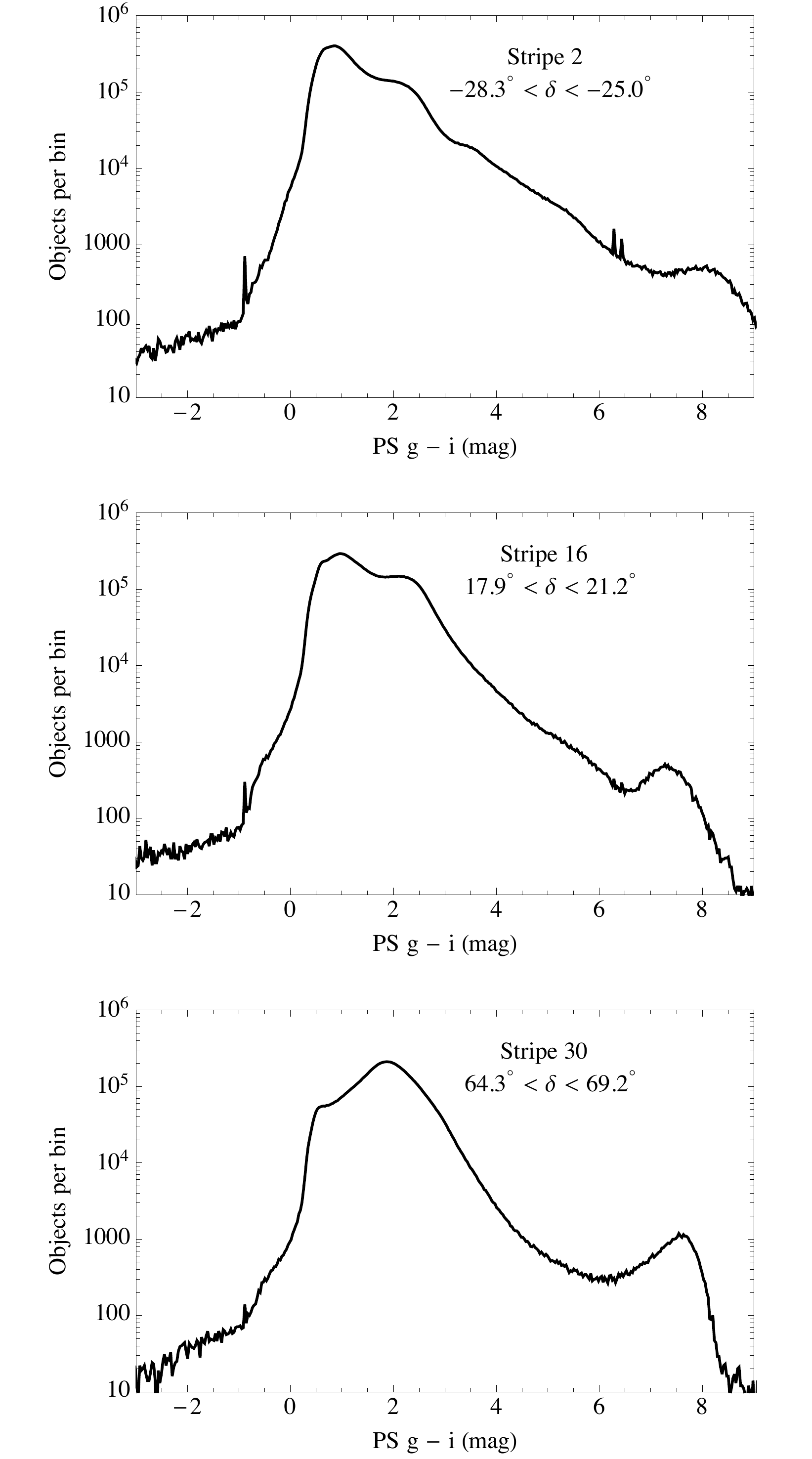}
\caption{Plot of the number distribution of reference objects on a log scale as a function of PS1 $g-i$ color in three \decl\ stripes.
The bin size is 0.03 mag.
}
\label{fig:ndistr}
\end{figure}

Figure~\ref{fig:ndistr} plots the number distribution of reference objects by color for Stripes 2, 16, and 30.
Stripe 2 is about $47^\circ$ in \decl\ south of the zenith, while Stripe 30 is about  $47^\circ$ in \decl\ north of the zenith.
Stripe 16 crosses the zenith.
More than 98\% of the reference objects lie in the color interval $0 < g-i < 4.5$.  There is a more rapid decline in the blue end end than in the red end.

\begin{figure*}
\centering
\includegraphics[width=\textwidth]{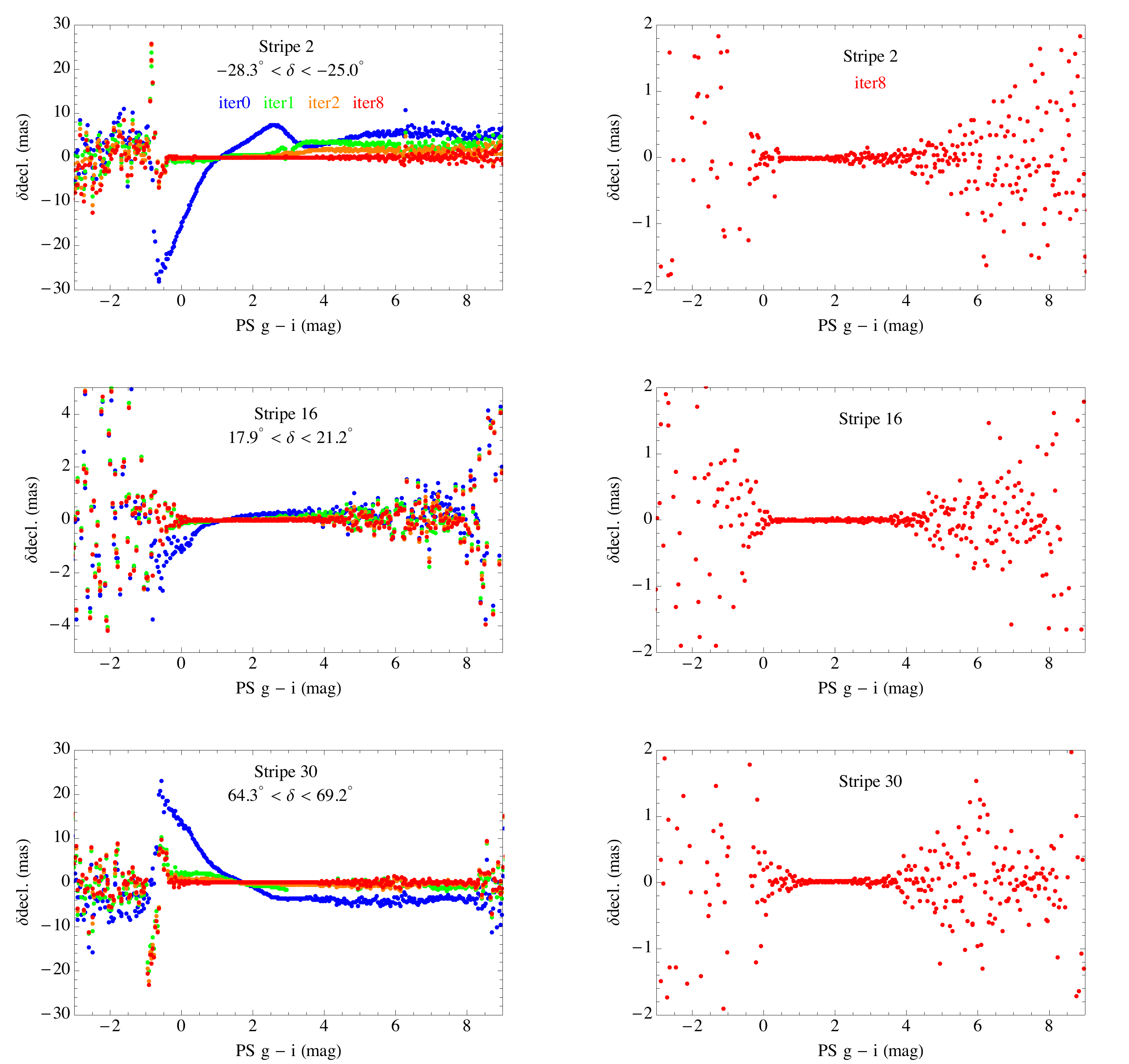}
\caption{Convergence of the iterative DCR correction.  The left column plots the Gaia EDR3 minus PS1 \decl\ residuals \deltadecl\ versus PS1 $g-i$ color in Stripes 2, 16, and 30 for iteration 0 (blue), iteration 1 (green), iteration 2 (orange), and iteration 8 (red). Notice that the scale is smaller for Stripe 16 in the left column.
The right column is a more detailed view of \deltadecl\  for  iteration 8.
}
\label{fig:iter}
\end{figure*}

The left column of Figure~\ref{fig:iter} plots the Gaia EDR3 minus PS1 \decl\ residuals as a function of PS1 $g-i$ color
that result from applying our algorithm.
The results are plotted for iterations 0, 1, 2, and 8. The right column is a more detailed view of the residuals for iteration 8.
As expected, the initial \decl\ residuals in iteration 0 (before DCR correction) are much smaller for Stripe 16 because fields in that region are generally observed closer to the zenith and have positions that are only weakly affected by refraction.\footnote{
In fact, due to the design of the PS1 telescope mount, the telescope cannot 
track objects closer than 10--20 degrees from the zenith. Sources
in Stripe 16 are consequently observed off the meridian, leading
to DCR shifts that affect \RA\ rather than decl.  Our algorithm
does not correct these smaller DCR offsets. See section~\ref{sec:errors} for more details.}
The initial (iteration 0) residuals in Stripes 2 and 30 are comparable in absolute value as would be expected for refraction by similar amounts.
These residuals are of opposite sign, as is also expected by the effects of refraction north and south of the zenith.

In the color interval  $0 < g-i < 4.5$ that contains more than 98\% of the objects, the iteration 0 residuals in Stripes 2 and 30 reach about 20 mas  at the blue end, but
are generally reduced well below 1 mas by iteration 8. Outside this color
interval, there is considerable scatter even for iteration 8, especially for the blue end where $g-i < -0.5$.
We investigated whether the
large scatter in the blue end could be due to stellar variability.
The PS1 observations of the same object in different filters were typically separated by months to years,
while they were nearly simultaneous with Gaia.
Therefore a given variable PS1 object could have errors in colors due to the magnitude changes in the different filters measured at different epochs. Such errors should not occur
with Gaia colors.  It is also possible that the scatter is due to other inaccuracies
in PS1 photometry.
To test these possibilities, we applied the correction algorithm using  Gaia $b-r$ colors instead of PS1 $g-i$ and obtained
a roughly similar level of scatter at the blue end where $b-r <0$.
Note that our algorithm relies on PS1 colors rather than Gaia colors
for the correction because we need to compute corrections for PS1
objects that do not have Gaia matches.

That suggests that variability is a less likely explanation than other sources of scatter in the PS1 colors.
We have found that extremely blue objects with $g-i < -0.5$ are
concentrated heavily in the middle of the Galactic plane, where
crowding makes their photometry, astrometry and colors all unreliable.
Even ``normal'' blue objects with $0 < g-i < 1$ are rare in this
region due to heavy dust reddening.  The DCR correction (and
indeed all properties) of objects with those extreme colors should
be considered highly uncertain. That is very likely to be the
source of the large scatter in the mean DCR correction for the
bluest objects.  We are investigating whether the catalog data for
these objects might be improved in a future reprocessing of the PS1
survey data, but for now users of PS1 DR2 should be skeptical of
objects in the crowded Galactic plane that are outliers in their
colors or other properties.

\begin{figure}
\includegraphics[width=\columnwidth]{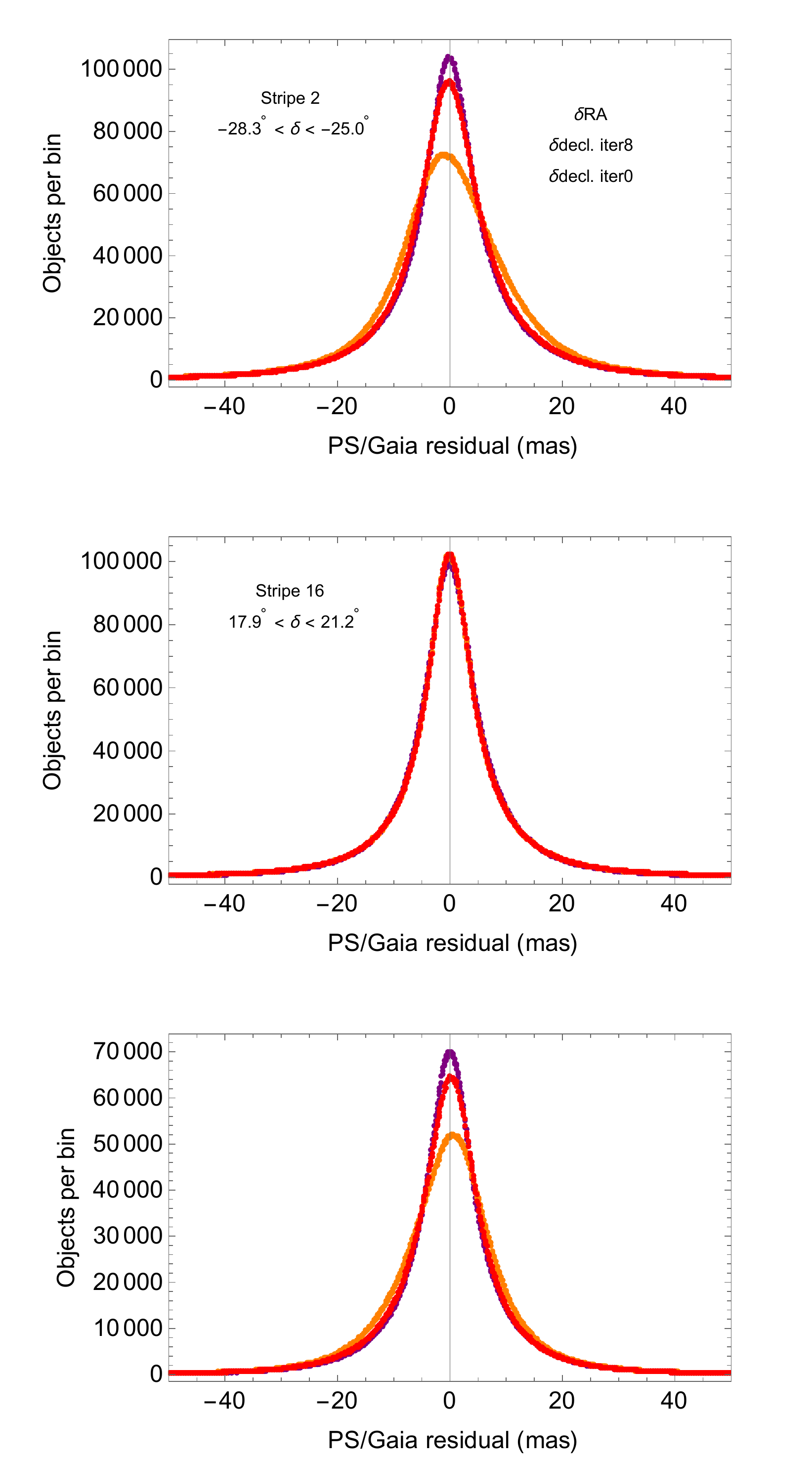}
\caption{Plot of the distributions of Gaia EDR3 minus PS1 residuals \deltaRA\ (purple),  \deltadecl\ for iteration 8 (red), and for iteration 0 (orange) in Stripes 2, 16, and 30. The bin size is 0.1 mas.  \RA\ residuals do not change across iterations.
The \RA\ and \decl\ distributions are much more similar after the DCR correction.
}
\label{fig:reshist}
\end{figure}

Figure~\ref{fig:reshist} compares the distributions of PS1/Gaia residuals for \RA\ and \decl\ at iterations 0 and 8. The DCR corrections lead to reductions in the \decl\ residuals
from iteration 0 (orange lines) to iteration 8 (red lines). Ideally, the residual distributions of \RA\ (purple lines) and \decl\
would be identical. But due to DCR effects that are stronger
in \decl\ than \RA, their distributions differ significantly before the DCR correction.
In the case of Stripe 16, all three distributions overlap
to the extent that they are indistinguishable on the plot because that stripe is only very slightly affected by refraction.
In Stripes 2 and 30, we see that the \RA\ residual distribution is much more peaked than the  \decl\ for iteration 0, where no DCR corrections have been made.
For iteration 8 the \decl\ distribution is much closer to the \RA\ distribution.
For Stripe 2 the ratio of the peaks in the \decl\ distribution to the \RA\ distribution for iteration 0 is 0.70 and for iteration 8 is 0.93.
For Stripe 30 the ratio of the peaks in the \decl\ distribution  to the \RA\ distribution  for iteration 0 is 0.73 and for iteration 8 is 0.92.
So the difference between the peaks is reduced by more than a factor of 3 from iteration 0  to iteration 8.

It is uncertain what is responsible for the remaining $\sim$8\% difference between the peaks of the \RA\ and \decl\ distributions
for iteration 8. One possibility is that it is due to fields being observed off the meridian. Fields observed on the meridian have
DCR shifts that are purely north-south with an amplitude determined by the object's color.  Fields observed off the meridian have shifts in both \decl\ and \RA\ with relative amplitudes that depend on the hour angle of the observation.
Due to this effect, there are additional systematic
DCR position shifts as a function of parallactic angle at the time of observation that we do not take into account. In principle such corrections could be made, and they were
made in the initial processing by \cite{Magnier2020}.  But we are using mean positions in making the PS1 astrometry corrections rather than the individual multi-epoch measurements; consequently our algorithm is limited to using the mean object properties rather
than adjusting the astrometry for every measurement independently.
The decision not to use the individual hour angles for every PS1 detection was based on our assessment that the much greater computational effort involved would not lead to much additional improvement in the astrometric accuracy.
We provide quantitative analysis in support of this choice in section~\ref{sec:errors}.

Another possible explanation is that inaccuracies in the PS1 colors lead to scatter in the applied DCR corrections.
As was mentioned above,
we obtained nearly identical results using the using Gaia $b-r$ color instead of PS1 $g-i$. The lack of improvement using
the more reliable Gaia colors makes this explanation less plausible. In particular,
for Stripe 30 after DCR corrections using Gaia colors, the ratio of the peaks in the \decl\ distribution  to the \RA\ distribution  for iteration 0 is 0.74 and for iteration 8 is 0.91.
So errors in the PS1 colors are not likely to affect the overall distribution of residuals much (although they can certainly affect individual objects).

\begin{figure}
\includegraphics[width=\columnwidth]{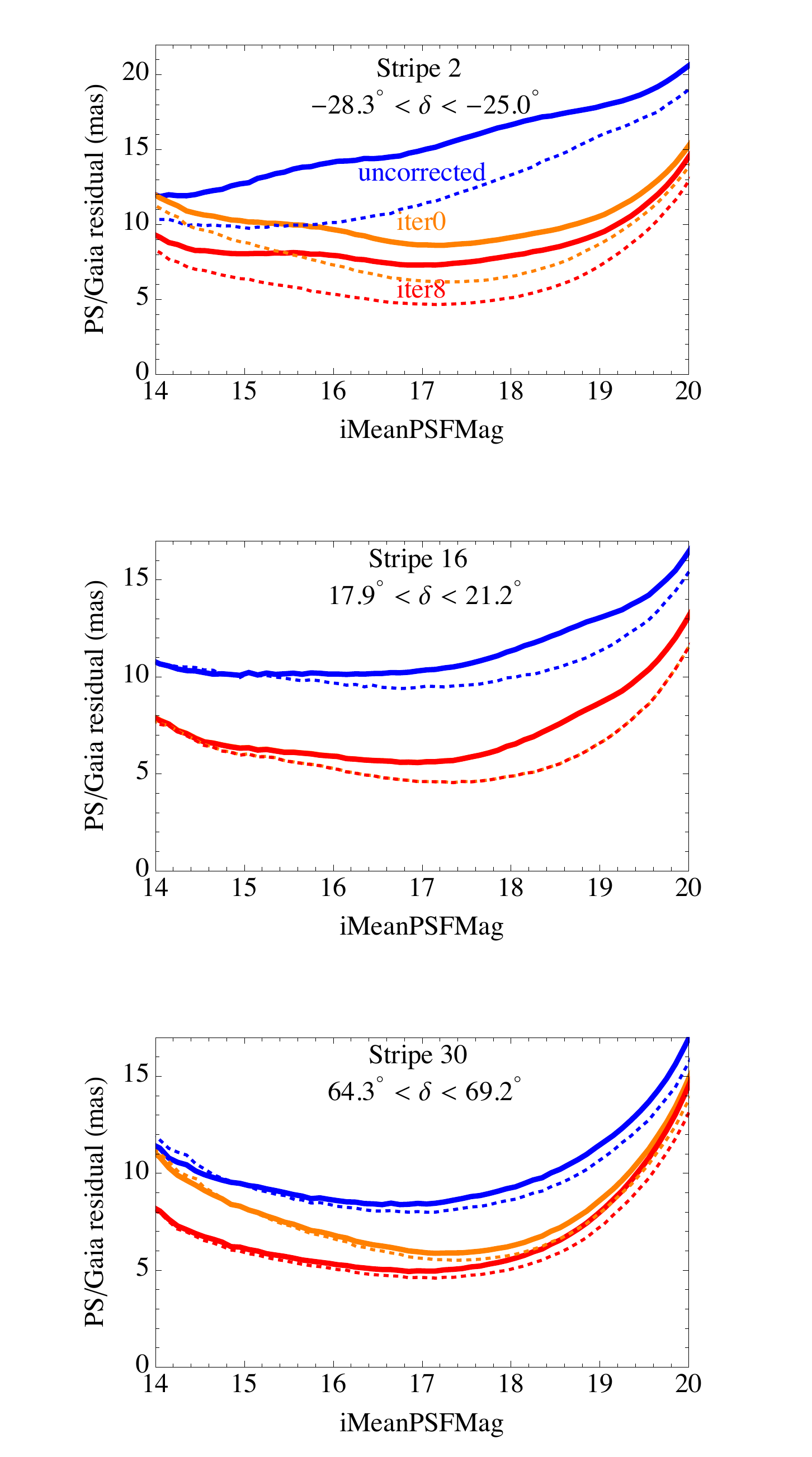}
\caption{Plot of the PS1/Gaia EDR3 residuals of reference objects as a  function of PS1 $i$-band magnitude for uncorrected (blue), iteration 0 (orange), and iteration 8 (red) cases, in Stripes 2, 16, and 30.
(The orange line is hidden behind the red line for Stripe 16.)
The solid lines  are for all reference objects in the stripe, while the dotted lines are for the very point-like reference objects having a point-source score greater than 0.99 \citep{Tachibana2018}.  }
\label{fig:mag}
\end{figure}

Figure~\ref{fig:mag}  plots the PS1/Gaia EDR3 residuals of reference objects for the uncorrected
(blue, no median neighbor shift corrections and no DCR corrections),
iteration 0 (orange, median neighbor shift corrections but no DCR corrections), and iteration 8 (red, median neighbor shift corrections and DCR corrections) cases as a function of PS1 $i$-band magnitude.
Iteration 0 applies the corrections from \citetalias{Lubow2021}, resulting in changes from the blue to the orange lines.
The DCR corrections result in the changes from the orange to the red lines.
The solid lines are for all reference
objects in the stripe for uncorrected, iteration 0, and iteration 8 cases, while
the dotted lines are for reference objects that are more point-like,
with a point-source score greater than 0.99 \citep{Tachibana2018} for the corresponding cases.

In most cases, the residuals in Figure~\ref{fig:mag} are minimized at an intermediate magnitude of about
$i=17\,\hbox{mag}$, as discussed in \citetalias{Lubow2021}. The  residual increase at the bright end is likely due to effects of saturation
(for the brightest stars) and to the Koppenh\"ofer Effect \citep{Magnier2020}, which generates brightness-dependent position errors in the PS1 detectors.
The increase at the faint end is due to the decrease in the signal-to-noise ratio.

In interpreting Figure~\ref{fig:mag}, we should keep in mind that Stripe 2 covers a region near the Galactic center, resulting in the most crowding of the three plotted stripes, and is well off the zenith, resulting in DCR effects. Stripe 16
passes through the zenith and has the smallest DCR effects.
Stripe 30 is the least crowded but is well off the zenith, resulting in DCR effects.
The residuals for the uncorrected case in Stripe 2  (solid blue line) generally increase monotonically with
magnitude.
The iteration 0 corrections do not provide much reduction of residuals for the brightest
objects in Stripe 2, but generally provide a substantial reduction for fainter objects.

Of the three stripes, Stripe 2 shows the largest reduction of residuals for very point-like objects (dotted lines)
compared to the general cases (solid lines) in the uncorrected, iteration 0,
and iteration 8 cases. This larger reduction is likely due to the effects
of crowding in Stripe 2.
Many of the objects with lower point-source scores are blended stars in crowded regions of the Galactic
halo and plane. A tighter point-source restriction excludes some objects where the PS1 astrometry is affected by
blending, leading to better astrometric residuals.
Stripe 30 does not show much astrometric improvement for the very point-like sources
(small differences between the solid and dashed lines of the same color)
because it is located far from the Galactic plane and is affected less by crowding.
The distributions
for the very point-like reference objects (dotted lines) in Stripe 2 are much more similar to the corresponding distributions 
in Stripe 30 than the less point-like cases (solid lines).
For Stripe 2 objects at 17 mag, the uncorrected residuals (solid blue line)
decrease by  about a factor of 3 for the DCR corrected, very point-like sources
(red dashed-line).

Since the objects in Stripe 16 experience very small DCR effects, this stripe allows us to
isolate the effects of using very point-like objects from the effects of DCR.
In  Stripe 16 the orange and red solid lines overlap
and the orange and red dotted lines overlap because the effects of refraction are weak.
Stripe 16 shows little change across the red and orange curves, both solid and dashed, in  Figure~\ref{fig:mag} for the brightest objects. For fainter objects, there is
a noticeable change  between the general and very point-like objects (solid versus dotted lines).
More point-like objects provide a greater advantage in astrometric accuracy for the fainter
objects, both because fainter objects are more likely to be blended with similar brightness
neighbors and also because the lower signal-to-noise ratio for fainter objects can make it
more difficult to confidently identify point-like objects for the reference sample.

For Stripes 2 and 30,  the DCR corrections are most effective for brighter objects.
The largest improvements are for the bright end of the plots for
Stripes  2 and 30 where the residuals are reduced by more than 25\%.
The DCR correction is less dramatic for fainter objects due to their
larger random astrometric errors
from noise. Random errors are not reduced by either the median shift correction or the DCR correction,
so the removal of those systematic errors has a smaller apparent effect on the
overall astrometric residuals for noisy, fainter sources.
There may also be a contribution from changing
color distributions with magnitude if 
the blue objects, which experience stronger DCR effects, are generally brighter than the red objects.

\begin{figure}
\includegraphics[width=\columnwidth]{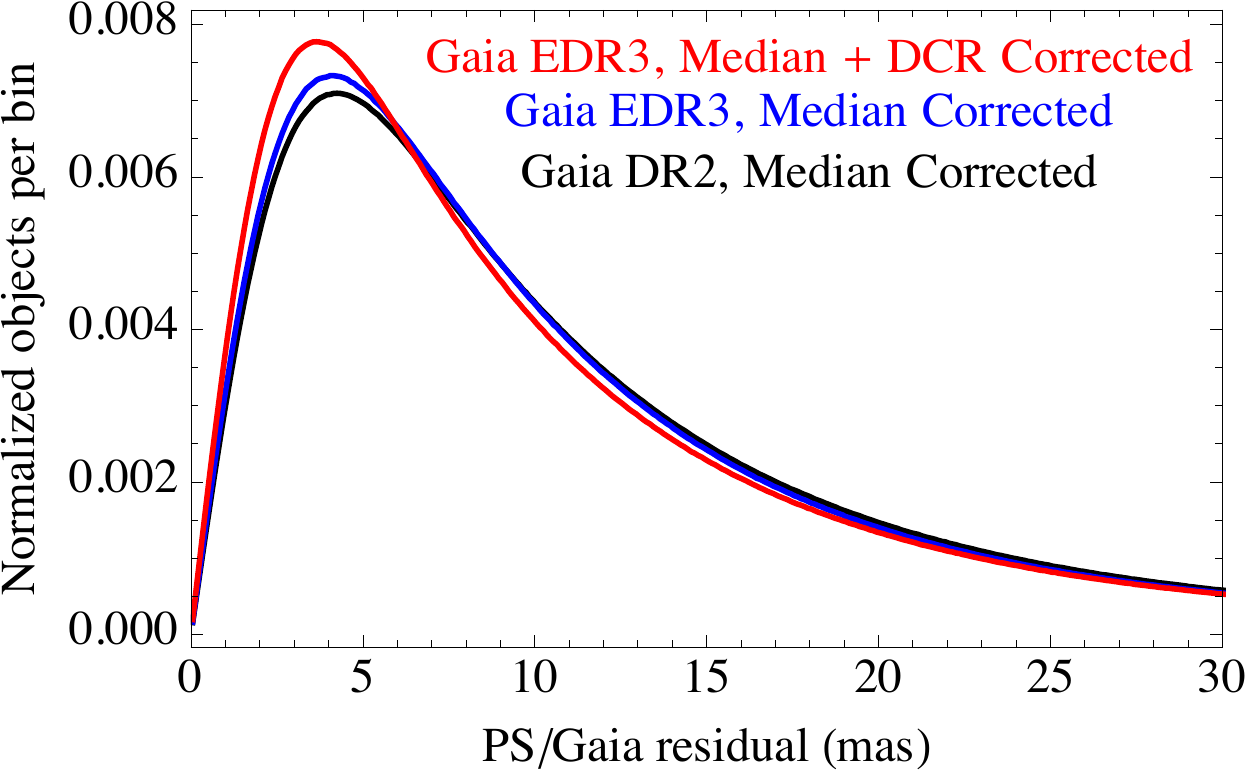}
\caption{Plot of the normalized number distribution of Gaia EDR3/PS1 positional residuals of reference objects after median neighbor shift correction algorithm using Gaia DR2 (black),
Gaia EDR3 (blue; iteration 0), and with additional DCR corrections (red; iteration 8) for all stripes.  The bin size is 0.1 mas. The number distributions are normalized such that their integrals are unity. }
\label{fig:globald}
\end{figure}

Figure~\ref{fig:globald} plots the normalized distribution functions of the distance between Gaia and PS1 cross matched objects for the cases of Gaia DR2, Gaia EDR3, and Gaia EDR3 with DCR corrections for all stripes.
The normalization is such that the areas under the curves are the same and equal to unity. The Gaia DR2 case is the same as in \citetalias{Lubow2021}.
As described by equation 3 of \citetalias{Lubow2021}, these functions are approximately of the form
of the Rayleigh distribution. As in \citetalias{Lubow2021}, the main tail of the plotted distributions from 20 mas to 40 mas are well fit to an exponential, while far into the tail (40 to 100 mas) the distributions follow a power law with index $\sim -2.8$.
We see that there is a small difference between the results for Gaia DR2 and Gaia EDR3 without DCR
corrections. Gaia EDR3 contained improvements to the accuracy of the astrometry, particularly in the accuracy of proper motions.  However, the epoch difference between PS1 and Gaia is  typically only about three years and PS1 is limited by its
ground-based resolution. Consequently,
these improvements produced a small reduction on residuals whose median values changed from 9.0 to 8.7 mas.
The unnormalized number distribution (not shown) reveals
an additional improvement in the peak of the distribution in using Gaia EDR3 over using Gaia DR2 because about 4\% more PS1 objects cross matched to Gaia EDR3.

\begin{figure}
\includegraphics[width=\columnwidth]{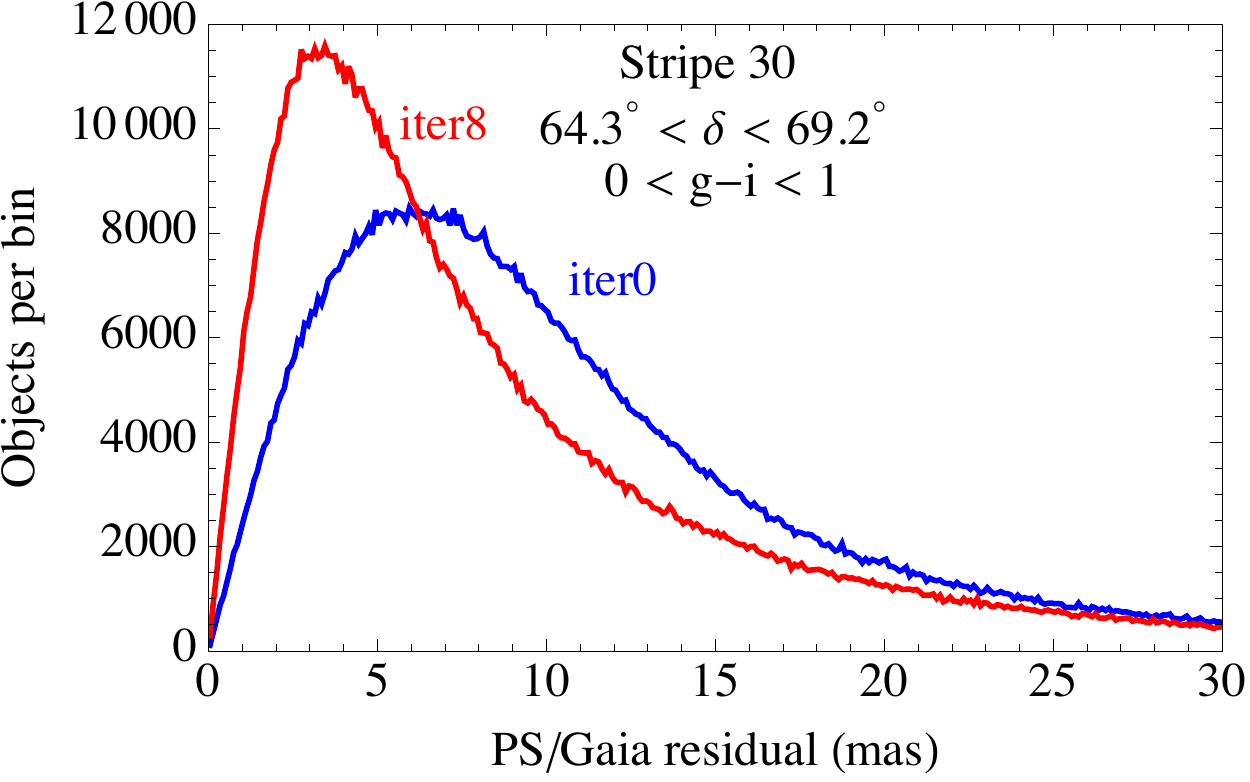}
\caption{Plot of the number distribution of Gaia EDR3/PS1 positional residuals of reference objects in Stripe 30 with colors $0<g-i<1$ for iteration 0 (blue) and iteration 8 (red). The bin size is 0.1 mas.}
\label{fig:str30bl}
\end{figure}

The DCR corrections reduced the overall median residual of reference objects to 8.3 mas, a reduction of about 8\% relative to Gaia DR2.
However, as we see in Figure~\ref{fig:iter} the DCR improvements can be much larger for blue objects observed away from the zenith.
For example, the median residuals in Stripe 30 for $0 < g-i <1$ decrease by 27\%, as seen in Figure~\ref{fig:str30bl}.

\section{Residual Errors in the Positions}
\label{sec:errors}

The algorithm we have adopted makes some simplifying assumptions.
One major assumption is that the DCR correction is a function solely
of the object's color and its decl.  That is reasonable
as long as fields are usually observed as they transit the meridian.
In this section we explore the accuracy of that assumption using information
on the actual distribution of PS1 observations, which is available
in the PS1 database.  We also measure the residual bias
in the DCR-corrected catalog by averaging the positional errors (compared
with Gaia EDR3) over small sky regions.
We conclude that despite the absence of any DCR correction in \RA, the
residual bias in the catalog positions is less than 1~mas in both \RA\ and \decl,
even for objects with very red or blue colors.

We have already seen in Figure~\ref{fig:mag} that Stripe 16 that contains the zenith involves 
small (less than 1 mas) DCR corrections in declination, since the orange and red curves
overlap. In addition, we see in Figure~\ref{fig:reshist} that the R.A. and decl. residual distributions are nearly identical. This suggests there are not significant errors in R.A.  due to DCR.

\subsection{Hour-angle distribution}
\label{sec:hourangle}

\begin{figure*}
\centering
\includegraphics[width=\textwidth]{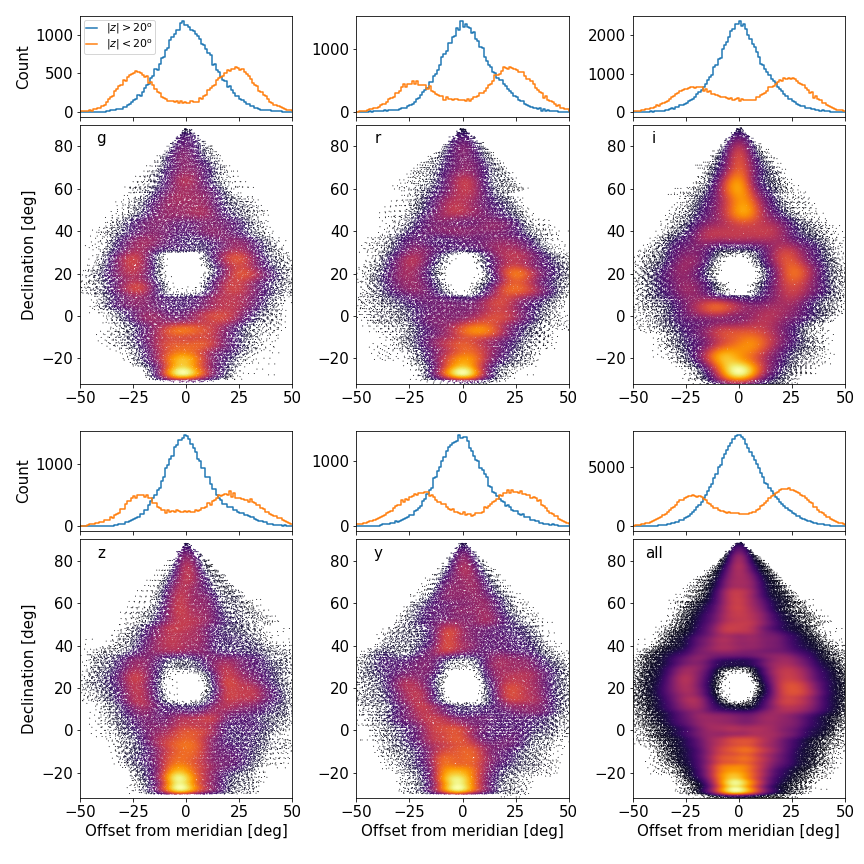}
\caption{Distribution of hour angles for PS1 observations.
The distributions are shown for each of the five PS1 filters
(\textit{grizy}) as well as the combined distribution for all filters.
The observations are sufficiently close to the 
meridian to produce DCR effects that are strongly concentrated along \decl\
except near the zenith (where the airmass of the observations are
always moderate, leading to small DCR effects).  The observations
that are off the meridian are approximately symmetrical about the
$y$-axis, leading to small biases ($<1$~mas) due to DCR in the \RA\
direction.  See the discussion for additional description of the figure.
}
\label{fig:haplot}
\end{figure*}

Figure~\ref{fig:haplot} shows the distribution of hour angles for
all the PS1 observations.
The distributions are shown separately for each of the five PS1 filters
(\textit{grizy}) as well as for all filters combined.
The $x$-axis is the distance to the meridian, $H\cos{\delta}$,
where $H$ is the hour angle of the observation in degrees (which
is zero as the target crosses the meridian) and $\delta$ is the
decl.  The color plot shows the density of observations as a function
of \decl\ (the $y$-axis), with individual pointings in sparse regions
shown as black points.  The hole in the middle is the position of the
zenith at the observatory and is a zone of avoidance due to the telescope's 
Alt-Az mount.  The histogram shows the distribution
integrated over \decl, with separate lines for the declination
band within $20^\circ$ of the zenith (orange) and points outside
that region (blue).

Large deviations from hour angles near zero would lead to significant DCR
effects in the \RA\ direction.  Strong asymmetries (e.g., observing targets systematically
as they are rising or setting) could also create systematic errors from DCR in R.A.  These
plots show that the hour angle distribution is relatively tight except near the zenith,
with a half-width at half maximum of $11.5^\circ$.  The distribution is particularly
tight in the far north and far south, where the DCR effects are the largest, which also
helps reduce the amplitudes of significant errors in the DCR correction.

\begin{figure}
\centering
\includegraphics[width=\linewidth]{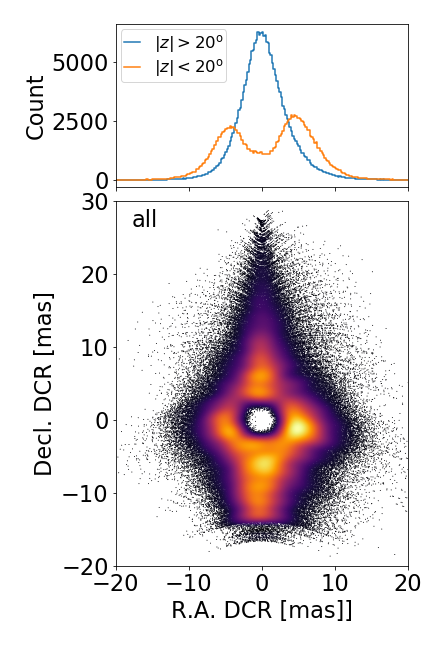}
\caption{Distribution of predicted \RA\ and \decl\ components of
DCR for PS1 observations of extremely blue stars.  The combined
hour-angle distribution (lower right panel of Fig.~\ref{fig:haplot})
was used, along with the largest amplitude correction that was
measured for any color stars in our sample.  The DCR effects in
\RA\ are typically small even in this worst-case example; the effects
on the mean positions are significantly smaller because most PS1
objects use many measurements to determine the means, reducing the
DCR bias by a factor $\sqrt{N}$ for $N$ observations (typically
$N=10$ to 60).
}
\label{fig:ra_dcr}
\end{figure}

How large are the errors that could be introduced by the observed
amount of hour-angle scatter?  As a partial answer to that question,
we have used the hour-angle data to construct a simple model of the
expected errors in R.A.  The hour angle and \decl\ of all
the observations from Figure~\ref{fig:haplot} have been used to
compute the direction of the DCR vector, which always points at the
zenith.  The DCR is assumed to scale with zenith distance $\zeta$ as
\begin{equation}
f_{\mathrm{DCR}}(\zeta) = B_{max}\, \tan{\zeta} / \tan\left(90^\circ-l_{obs}\right) \quad ,
\end{equation}
where $l_{obs} = 20.7082^\circ$ is the latitude of the observatory,
$90^\circ - l_{obs}$ is the largest zenith distance, and
$B_{max} = 30\,\mathrm{mas}$ is the largest DCR correction that
we compute for the most extreme (blue) stars.  Figure~\ref{fig:ra_dcr}
shows the resulting distribution of \RA\ and \decl\ DCR components;
the histogram again shows the distribution integrated over \decl,
with separate lines for the declination band within $20^\circ$ of
the zenith (orange) and points outside that region (blue).

Despite the choice of a worst-case blue color for the DCR scaling, the
half-width at half maximum of the \RA\ DCR distribution is only
2.7~mas.  The remaining DCR errors for the mean positions in the
catalog will be even smaller.  If there are $N$ observations with
hour angles sampled randomly from this distribution, the error in
the mean will be smaller by a factor $\sqrt{N}$.  PS1 objects can
have $N$ ranging from 3 (the lowest $N$ considered for this paper)
to more than 60, with typical objects having 10 or more measurements.
We can therefore expect biases in \RA\ from residual uncorrected DCR of
around 1~mas or less.

\begin{figure*}
\centering
\includegraphics[width=\textwidth]{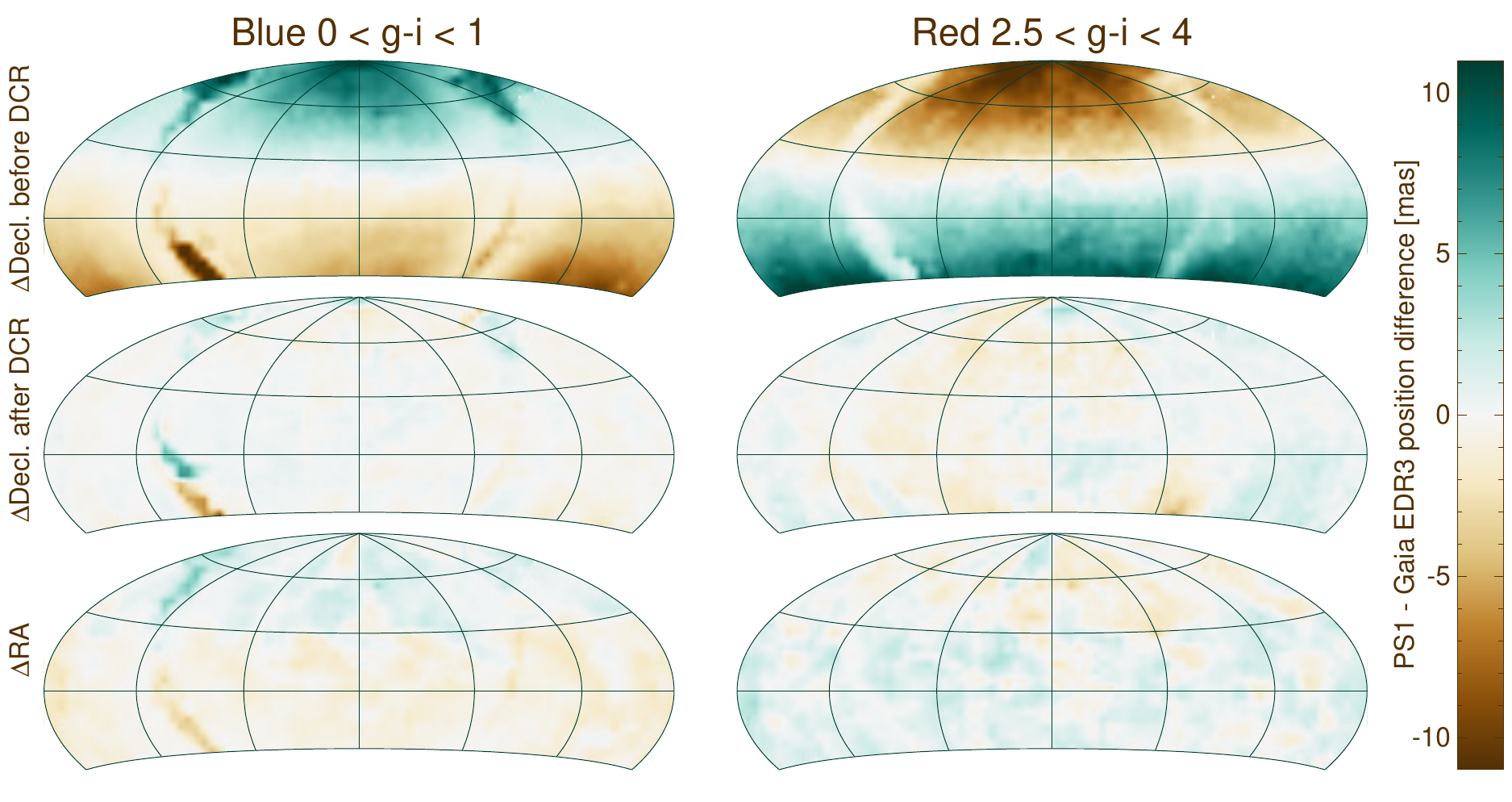}
\caption{Mean positional bias in PS1 catalog positions compared
with Gaia as a function of sky position for blue stars (left column)
and red stars (right column).  The top row shows the \decl\ offset
before the DCR correction (iteration 0), and the middle row shows
the \decl\ offset after the DCR correction (iteration 8).  The
bottom row shows the offset in \RA, which is not changed by the DCR
correction.  The same scale is used for all panels.  The faintly
seen difference between higher and lower declination regions in the
\RA\ plot is due to uncorrected DCR effects.  However, the very
small amplitude ($<1$~mas), combined with the complex nature of an
algorithm to correct it, led us not to attempt a DCR correction for
the R.A.
}
\label{fig:skybias}
\end{figure*}

\subsection{Measured bias in positions}
\label{sec:bias}

Finally, we directly measure the bias in our positions
by comparing the final corrected positions to Gaia EDR3.  Recall
that we explicitly avoid using the Gaia measurement of the object
itself in calculating the correction \citepalias{Lubow2021}, leaving
our PS1 positions for the PS1-Gaia reference sample uncontaminated
and independent of the Gaia measurement.  The average positional error
of a group of PS1-Gaia stars is, if there are enough stars, a measure of the
bias in the PS1 coordinate system.

We computed the mean offsets in \RA\ and \decl\ between the PS1 and
Gaia positions in small sky regions covering approximately
$4.5^\circ\times4.5^\circ$.  The regions were chosen to be large
enough to have a sufficient red and blue stars for an accurate
mean calculation (usually with $>1000$ objects per spatial bin),
while remaining compact enough to reveal any sky regions where the
mean error is unusually large.  Our calculation is similar to using
PS1 as an astrometric reference catalog for aligning an astronomical
image.

The results are shown in 
Figure~\ref{fig:skybias}, which shows the average positional bias as a function of sky positions
for blue stars ($0 < g-i < 1$, left column) and red stars ($2.5 < g-i < 4$, right column).
The \decl\ offset before the DCR correction (top row) is
clearly dominated by the large DCR effects, with the sign of the offsets changing
around the declination of the zenith, and with the red and blue stars having offsets
in opposite directions.  After the DCR correction (middle row), the \decl\ errors are
well behaved and small except for some local errors in the Galactic plane.

\begin{figure}
\includegraphics[width=\linewidth]{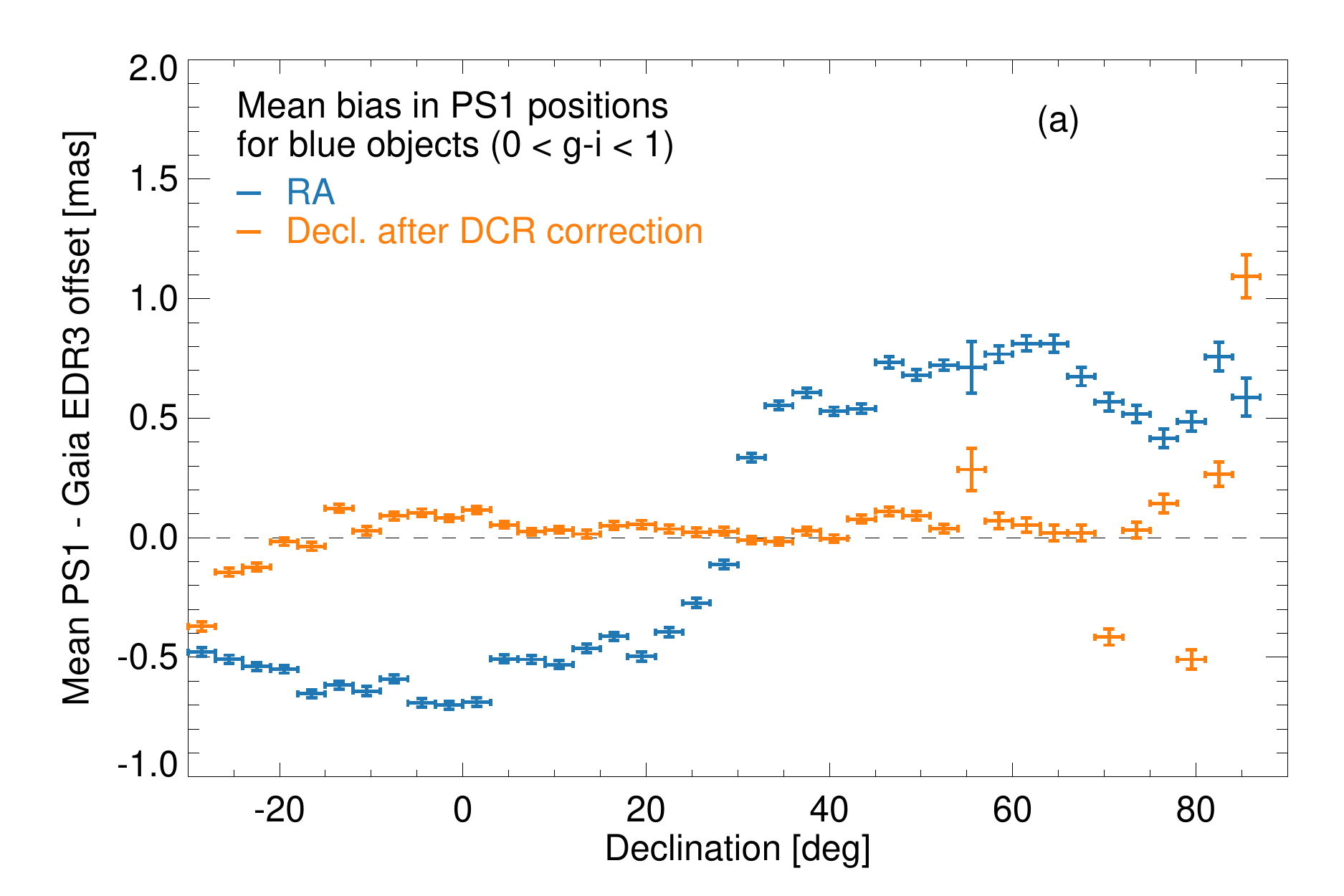}
\includegraphics[width=\linewidth]{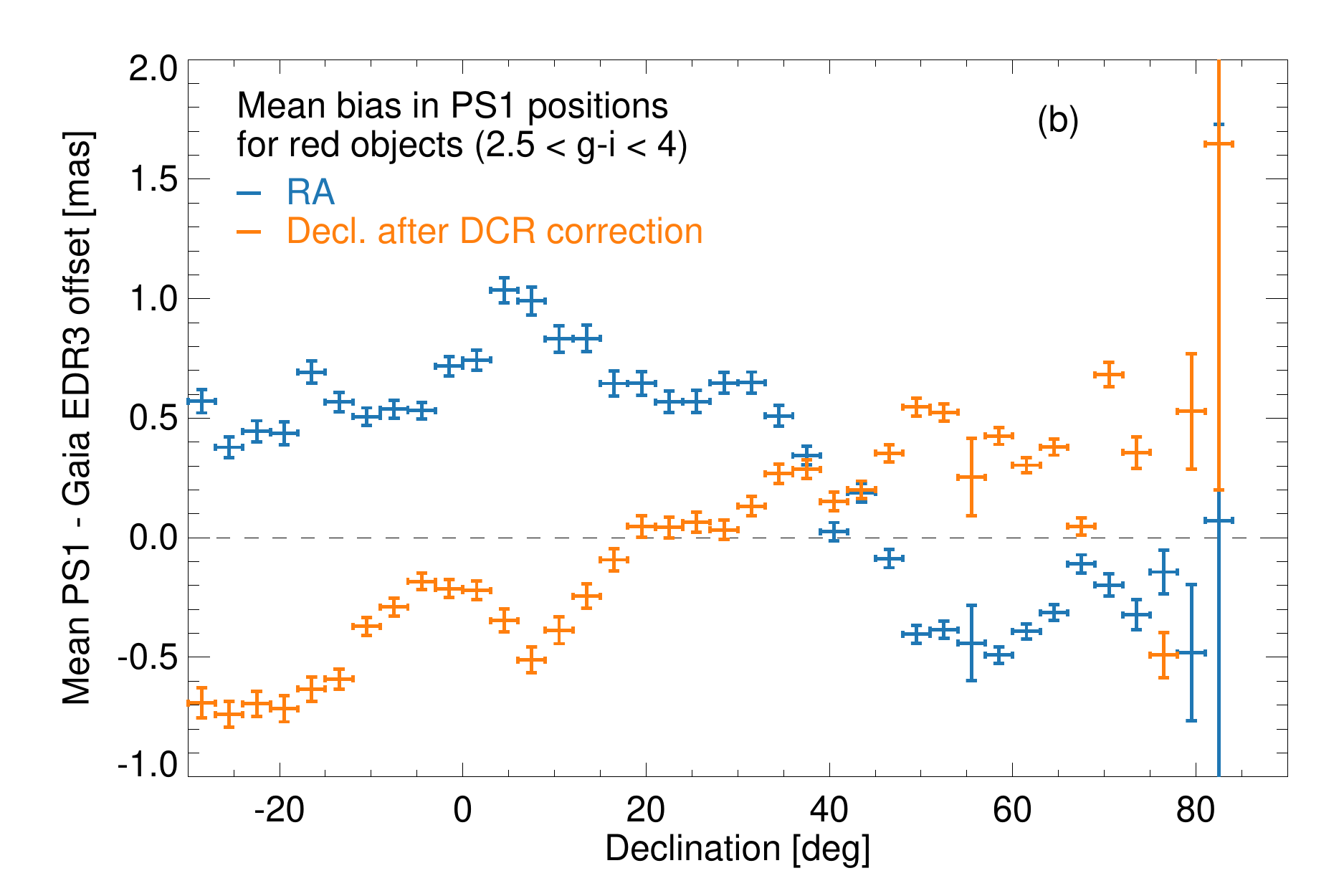}
\caption{Mean positional bias in PS1 catalog positions 
after DCR correction for (a) blue stars and (b) red stars.
The remaining bias in the coordinates is less than 1~mas, even
for these very red and blue objects.
}
\label{fig:bias}
\end{figure}

The structures in the Galactic plane, which are particularly visible
before the DCR correction, are the result of changes in the average
colors of reference stars near the plane (Fig.~\ref{fig:skycolor}).
In the center of the plane, dust absorption makes typical reference
stars as red as our ``red'' sample, meaning that there are no
DCR-induced offsets between them.  As a result, the \decl\ error
before DCR correction is close to zero for red objects (top right
panel).  Conversely, the (rare) blue stars in the plane have enormous
color differences compared with the red stars, leading to very large
DCR errors (top left panel).  After correction the errors for blue
stars are greatly reduced in the plane, though they still remain
at a low level.  That is at least partly attributable to severe
crowding that limits the quality of both astrometry and photometry
in the plane.  There are in any case few blue stars in the plane
that are affected by the remaining errors.

Errors in the \RA, which is not changed by our DCR correction algorithm,
are generally small and are comparable to the corrected \decl\ errors
(as expected from Fig.~\ref{fig:reshist}).  There is no evidence for of any
seasonal variation in the \RA\ errors, which could have resulted if there
were changes in the observing strategy due to different weather patterns
at different times of the year.

A slight difference between higher and lower declination regions
is visible in the \RA\ plot.  That difference changes sign for red
and blue objects, indicating that it is almost certainly an uncorrected
DCR effect.  However, its amplitude is very small.
Since the \RA\
bias is primarily a function of \decl, Figure~\ref{fig:bias} shows
the mean error for the post-DCR-correction errors as a function of
decl.  While both the red and blue stars show the characteristic
sign change around the \decl\ of the zenith that is expected from
DCR, the amplitude of the remaining bias in both \RA\ and \decl\
is less than 1~mas, even for these stars having extreme colors.
For stars with more typical colors, the biases are much smaller.
That makes our improved PS1 positions useful as an astrometric
reference catalog for applications that require fainter objects
than are found in the Gaia catalog.

\section{Summary}
\label{sec:summary}

We extended the astrometric correction algorithm of \citetalias{Lubow2021} to use
Gaia EDR3 and include corrections for differential chromatic
refraction (DCR).  Gaia EDR3 provides an improvement of $\sim 3\%$
in PS1 astrometric accuracy of all reference objects compared with
Gaia DR2.  DCR corrections provide an additional improvement of
$\sim 5\%$ (Fig.~\ref{fig:globald}).  However, the DCR corrections
improve the astrometric accuracy as much as $\sim 30\%$ for brighter
objects ($i < 17$) or bluer objects that are well off the zenith
(Figs.~\ref{fig:mag} and \ref{fig:str30bl}).  The amplitude of
systematic astrometric errors from these effects is substantially
reduced to less than 1~mas for objects with PS1 colors in the range
$0 < g-i < 4.5$ (Fig.~\ref{fig:bias}), which makes this a useful
astrometric reference catalog in fields where there are few Gaia
stars available.
As a result of the DCR corrections, the \RA\ and \decl\ residual
distributions are quite similar (Fig.~\ref{fig:reshist}).  We are
making these improvements to astrometry available soon in the SQL
query interface called CasJobs at
\url{https://mastweb.stsci.edu/ps1casjobs} and through the Barbara A.\ Mikulski
Archive for Space Telescopes (MAST) 
catalogs interface at \url{https://catalogs.mast.stsci.edu}.

\section*{Acknowledgments}

We thank Mike Fall and the Astrometry Working Group at STScI for
motivating this work and providing feedback during the preliminary phase.
In particular, we thank Stefano Casertano for examining the level of improvement
that could be possible for PS1 astrometry by using Gaia.
We also thank
Ciprian Berghia, Valeri Makarov, and Juliene Frouard for insightful discussions
about the limitations on the use of Gaia in the current PS1 astrometry.

The Pan-STARRS1 Surveys (PS1) and the PS1 public science archive have been made possible through contributions by the Institute for Astronomy, the University of Hawaii, the Pan-STARRS Project Office, the Max-Planck Society and its participating institutes, the Max Planck Institute for Astronomy, Heidelberg and the Max Planck Institute for Extraterrestrial Physics, Garching, The Johns Hopkins University, Durham University, the University of Edinburgh, the Queen's University Belfast, the Harvard-Smithsonian Center for Astrophysics, the Las Cumbres Observatory Global Telescope Network Incorporated, the National Central University of Taiwan, the Space Telescope Science Institute, the National Aeronautics and Space Administration under Grant No. NNX08AR22G issued through the Planetary Science Division of the NASA Science Mission Directorate, the National Science Foundation Grant No. AST-1238877, the University of Maryland, Eotvos Lorand University (ELTE), the Los Alamos National Laboratory, and the Gordon and Betty Moore Foundation.

Data presented in this paper were obtained from
the Barbara A.\ Mikulski Archive for Space Telescopes (MAST) at the Space
Telescope Science Institute. The specific observations analyzed can
be accessed via
\dataset[doi:10.17909/s0zg-jx37]{https://doi.org/10.17909/s0zg-jx37}.

This work has made use of data from the European Space Agency (ESA) mission
{\it Gaia} (\url{https://www.cosmos.esa.int/gaia}), processed by the {\it Gaia}
Data Processing and Analysis Consortium (DPAC,
\url{https://www.cosmos.esa.int/web/gaia/dpac/consortium}). Funding for the DPAC
has been provided by national institutions, in particular the institutions
participating in the {\it Gaia} Multilateral Agreement.  We used the Gaia EDR3
catalog, which can be accessed via 
\dataset[doi:10.5270/esa-1ugzkg7]{https://doi.org/10.5270/esa-1ugzkg7}.

\facilities{MAST (Pan-STARRS), Gaia}


\bibliographystyle{aasjournal}
\bibliography{references}

\begin{thebibliography}{}
\expandafter\ifx\csname natexlab\endcsname\relax\def\natexlab#1{#1}\fi
\providecommand{\url}[1]{\href{#1}{#1}}
\providecommand{\dodoi}[1]{doi:~\href{http://doi.org/#1}{\nolinkurl{#1}}}
\providecommand{\doeprint}[1]{\href{http://ascl.net/#1}{\nolinkurl{http://ascl.net/#1}}}
\providecommand{\doarXiv}[1]{\href{https://arxiv.org/abs/#1}{\nolinkurl{https://arxiv.org/abs/#1}}}

\bibitem[{{Budav{\'a}ri} {et~al.}(2010){Budav{\'a}ri}, {Szalay}, \&
  {Fekete}}]{Budavari2010}
{Budav{\'a}ri}, T., {Szalay}, A.~S., \& {Fekete}, G. 2010, \pasp, 122, 1375,
  \dodoi{10.1086/657302}

\bibitem[{{Chambers} {et~al.}(2016){Chambers}, {Magnier}, {Metcalfe},
  {Flewelling}, {Huber}, {Waters}, {Denneau}, {Draper}, {Farrow}, {Finkbeiner},
  {Holmberg}, {Koppenhoefer}, {Price}, {Rest}, {Saglia}, {Schlafly}, {Smartt},
  {Sweeney}, {Wainscoat}, {Burgett}, {Chastel}, {Grav}, {Heasley}, {Hodapp},
  {Jedicke}, {Kaiser}, {Kudritzki}, {Luppino}, {Lupton}, {Monet}, {Morgan},
  {Onaka}, {Shiao}, {Stubbs}, {Tonry}, {White}, {Ba{\~n}ados}, {Bell},
  {Bender}, {Bernard}, {Boegner}, {Boffi}, {Botticella}, {Calamida},
  {Casertano}, {Chen}, {Chen}, {Cole}, {Deacon}, {Frenk}, {Fitzsimmons},
  {Gezari}, {Gibbs}, {Goessl}, {Goggia}, {Gourgue}, {Goldman}, {Grant},
  {Grebel}, {Hambly}, {Hasinger}, {Heavens}, {Heckman}, {Henderson}, {Henning},
  {Holman}, {Hopp}, {Ip}, {Isani}, {Jackson}, {Keyes}, {Koekemoer}, {Kotak},
  {Le}, {Liska}, {Long}, {Lucey}, {Liu}, {Martin}, {Masci}, {McLean}, {Mindel},
  {Misra}, {Morganson}, {Murphy}, {Obaika}, {Narayan}, {Nieto-Santisteban},
  {Norberg}, {Peacock}, {Pier}, {Postman}, {Primak}, {Rae}, {Rai}, {Riess},
  {Riffeser}, {Rix}, {R{\"o}ser}, {Russel}, {Rutz}, {Schilbach}, {Schultz},
  {Scolnic}, {Strolger}, {Szalay}, {Seitz}, {Small}, {Smith}, {Soderblom},
  {Taylor}, {Thomson}, {Taylor}, {Thakar}, {Thiel}, {Thilker}, {Unger},
  {Urata}, {Valenti}, {Wagner}, {Walder}, {Walter}, {Watters}, {Werner},
  {Wood-Vasey}, \& {Wyse}}]{Chambers2016}
{Chambers}, K.~C., {Magnier}, E.~A., {Metcalfe}, N., {et~al.} 2016, arXiv
  e-prints, arXiv:1612.05560.
\newblock \doarXiv{1612.05560}

\bibitem[{{Flewelling} {et~al.}(2020){Flewelling}, {Magnier}, {Chambers},
  {Heasley}, {Holmberg}, {Huber}, {Sweeney}, {Waters}, {Calamida}, {Casertano},
  {Chen}, {Farrow}, {Hasinger}, {Henderson}, {Long}, {Metcalfe}, {Narayan},
  {Nieto-Santisteban}, {Norberg}, {Rest}, {Saglia}, {Szalay}, {Thakar},
  {Tonry}, {Valenti}, {Werner}, {White}, {Denneau}, {Draper}, {Hodapp},
  {Jedicke}, {Kaiser}, {Kudritzki}, {Price}, {Wainscoat}, {Chastel}, {McLean},
  {Postman}, \& {Shiao}}]{Flewelling2020}
{Flewelling}, H.~A., {Magnier}, E.~A., {Chambers}, K.~C., {et~al.} 2020, \apjs,
  251, 7, \dodoi{10.3847/1538-4365/abb82d}

\bibitem[{{Gaia Collaboration} {et~al.}(2016){Gaia Collaboration}, {Prusti},
  {de Bruijne}, {Brown}, {Vallenari}, {Babusiaux}, {Bailer-Jones}, {Bastian},
  {Biermann}, {Evans}, {Eyer}, {Jansen}, {Jordi}, {Klioner}, {Lammers},
  {Lindegren}, {Luri}, {Mignard}, {Milligan}, {Panem}, {Poinsignon},
  {Pourbaix}, {Randich}, {Sarri}, {Sartoretti}, {Siddiqui}, {Soubiran},
  {Valette}, {van Leeuwen}, {Walton}, {Aerts}, {Arenou}, {Cropper}, {Drimmel},
  {H{\o}g}, {Katz}, {Lattanzi}, {O'Mullane}, {Grebel}, {Holland}, {Huc},
  {Passot}, {Bramante}, {Cacciari}, {Casta{\~n}eda}, {Chaoul}, {Cheek}, {De
  Angeli}, {Fabricius}, {Guerra}, {Hern{\'a}ndez}, {Jean-Antoine-Piccolo},
  {Masana}, {Messineo}, {Mowlavi}, {Nienartowicz}, {Ord{\'o}{\~n}ez-Blanco},
  {Panuzzo}, {Portell}, {Richards}, {Riello}, {Seabroke}, {Tanga},
  {Th{\'e}venin}, {Torra}, {Els}, {Gracia-Abril}, {Comoretto},
  {Garcia-Reinaldos}, {Lock}, {Mercier}, {Altmann}, {Andrae}, {Astraatmadja},
  {Bellas-Velidis}, {Benson}, {Berthier}, {Blomme}, {Busso}, {Carry},
  {Cellino}, {Clementini}, {Cowell}, {Creevey}, {Cuypers}, {Davidson}, {De
  Ridder}, {de Torres}, {Delchambre}, {Dell'Oro}, {Ducourant}, {Fr{\'e}mat},
  {Garc{\'\i}a-Torres}, {Gosset}, {Halbwachs}, {Hambly}, {Harrison}, {Hauser},
  {Hestroffer}, {Hodgkin}, {Huckle}, {Hutton}, {Jasniewicz}, {Jordan},
  {Kontizas}, {Korn}, {Lanzafame}, {Manteiga}, {Moitinho}, {Muinonen},
  {Osinde}, {Pancino}, {Pauwels}, {Petit}, {Recio-Blanco}, {Robin}, {Sarro},
  {Siopis}, {Smith}, {Smith}, {Sozzetti}, {Thuillot}, {van Reeven}, {Viala},
  {Abbas}, {Abreu Aramburu}, {Accart}, {Aguado}, {Allan}, {Allasia},
  {Altavilla}, {{\'A}lvarez}, {Alves}, {Anderson}, {Andrei}, {Anglada Varela},
  {Antiche}, {Antoja}, {Ant{\'o}n}, {Arcay}, {Atzei}, {Ayache}, {Bach},
  {Baker}, {Balaguer-N{\'u}{\~n}ez}, {Barache}, {Barata}, {Barbier}, {Barblan},
  {Baroni}, {Barrado y Navascu{\'e}s}, {Barros}, {Barstow}, {Becciani},
  {Bellazzini}, {Bellei}, {Bello Garc{\'\i}a}, {Belokurov}, {Bendjoya},
  {Berihuete}, {Bianchi}, {Bienaym{\'e}}, {Billebaud}, {Blagorodnova},
  {Blanco-Cuaresma}, {Boch}, {Bombrun}, {Borrachero}, {Bouquillon}, {Bourda},
  {Bouy}, {Bragaglia}, {Breddels}, {Brouillet}, {Br{\"u}semeister},
  {Bucciarelli}, {Budnik}, {Burgess}, {Burgon}, {Burlacu}, {Busonero}, {Buzzi},
  {Caffau}, {Cambras}, {Campbell}, {Cancelliere}, {Cantat-Gaudin}, {Carlucci},
  {Carrasco}, {Castellani}, {Charlot}, {Charnas}, {Charvet}, {Chassat},
  {Chiavassa}, {Clotet}, {Cocozza}, {Collins}, {Collins}, {Costigan}, {Crifo},
  {Cross}, {Crosta}, {Crowley}, {Dafonte}, {Damerdji}, {Dapergolas}, {David},
  {David}, {De Cat}, {de Felice}, {de Laverny}, {De Luise}, {De March}, {de
  Martino}, {de Souza}, {Debosscher}, {del Pozo}, {Delbo}, {Delgado},
  {Delgado}, {di Marco}, {Di Matteo}, {Diakite}, {Distefano}, {Dolding}, {Dos
  Anjos}, {Drazinos}, {Dur{\'a}n}, {Dzigan}, {Ecale}, {Edvardsson}, {Enke},
  {Erdmann}, {Escolar}, {Espina}, {Evans}, {Eynard Bontemps}, {Fabre},
  {Fabrizio}, {Faigler}, {Falc{\~a}o}, {Farr{\`a}s Casas}, {Faye}, {Federici},
  {Fedorets}, {Fern{\'a}ndez-Hern{\'a}ndez}, {Fernique}, {Fienga}, {Figueras},
  {Filippi}, {Findeisen}, {Fonti}, {Fouesneau}, {Fraile}, {Fraser}, {Fuchs},
  {Furnell}, {Gai}, {Galleti}, {Galluccio}, {Garabato}, {Garc{\'\i}a-Sedano},
  {Gar{\'e}}, {Garofalo}, {Garralda}, {Gavras}, {Gerssen}, {Geyer}, {Gilmore},
  {Girona}, {Giuffrida}, {Gomes}, {Gonz{\'a}lez-Marcos},
  {Gonz{\'a}lez-N{\'u}{\~n}ez}, {Gonz{\'a}lez-Vidal}, {Granvik}, {Guerrier},
  {Guillout}, {Guiraud}, {G{\'u}rpide}, {Guti{\'e}rrez-S{\'a}nchez}, {Guy},
  {Haigron}, {Hatzidimitriou}, {Haywood}, {Heiter}, {Helmi}, {Hobbs},
  {Hofmann}, {Holl}, {Holland }, {Hunt}, {Hypki}, {Icardi}, {Irwin}, {Jevardat
  de Fombelle}, {Jofr{\'e}}, {Jonker}, {Jorissen}, {Julbe}, {Karampelas},
  {Kochoska}, {Kohley}, {Kolenberg}, {Kontizas}, {Koposov}, {Kordopatis},
  {Koubsky}, {Kowalczyk}, {Krone-Martins}, {Kudryashova}, {Kull}, {Bachchan},
  {Lacoste-Seris}, {Lanza}, {Lavigne}, {Le Poncin-Lafitte}, {Lebreton},
  {Lebzelter}, {Leccia}, {Leclerc}, {Lecoeur-Taibi}, {Lemaitre}, {Lenhardt},
  {Leroux}, {Liao}, {Licata}, {Lindstr{\o}m}, {Lister}, {Livanou}, {Lobel},
  {L{\"o}ffler}, {L{\'o}pez}, {Lopez-Lozano}, {Lorenz}, {Loureiro},
  {MacDonald}, {Magalh{\~a}es Fernandes}, {Managau}, {Mann}, {Mantelet},
  {Marchal}, {Marchant}, {Marconi}, {Marie}, {Marinoni}, {Marrese},
  {Marschalk{\'o}}, {Marshall}, {Mart{\'\i}n-Fleitas}, {Martino}, {Mary},
  {Matijevi{\v{c}}}, {Mazeh}, {McMillan}, {Messina}, {Mestre}, {Michalik},
  {Millar}, {Miranda}, {Molina}, {Molinaro}, {Molinaro}, {Moln{\'a}r},
  {Moniez}, {Montegriffo}, {Monteiro}, {Mor}, {Mora}, {Morbidelli}, {Morel},
  {Morgenthaler}, {Morley}, {Morris}, {Mulone}, {Muraveva}, {Musella},
  {Narbonne}, {Nelemans}, {Nicastro}, {Noval}, {Ord{\'e}novic},
  {Ordieres-Mer{\'e}}, {Osborne}, {Pagani}, {Pagano}, {Pailler}, {Palacin},
  {Palaversa}, {Parsons}, {Paulsen}, {Pecoraro}, {Pedrosa}, {Pentik{\"a}inen},
  {Pereira}, {Pichon}, {Piersimoni}, {Pineau}, {Plachy}, {Plum}, {Poujoulet},
  {Pr{\v{s}}a}, {Pulone}, {Ragaini}, {Rago}, {Rambaux}, {Ramos-Lerate},
  {Ranalli}, {Rauw}, {Read}, {Regibo}, {Renk}, {Reyl{\'e}}, {Ribeiro},
  {Rimoldini}, {Ripepi}, {Riva}, {Rixon}, {Roelens}, {Romero-G{\'o}mez},
  {Rowell}, {Royer}, {Rudolph}, {Ruiz-Dern}, {Sadowski}, {Sagrist{\`a}
  Sell{\'e}s}, {Sahlmann}, {Salgado}, {Salguero}, {Sarasso}, {Savietto},
  {Schnorhk}, {Schultheis}, {Sciacca}, {Segol}, {Segovia}, {Segransan},
  {Serpell}, {Shih}, {Smareglia}, {Smart}, {Smith}, {Solano}, {Solitro},
  {Sordo}, {Soria Nieto}, {Souchay}, {Spagna}, {Spoto}, {Stampa}, {Steele},
  {Steidelm{\"u}ller}, {Stephenson}, {Stoev}, {Suess}, {S{\"u}veges}, {Surdej},
  {Szabados}, {Szegedi-Elek}, {Tapiador}, {Taris}, {Tauran}, {Taylor},
  {Teixeira}, {Terrett}, {Tingley}, {Trager}, {Turon}, {Ulla}, {Utrilla},
  {Valentini}, {van Elteren}, {Van Hemelryck}, {van Leeuwen}, {Varadi},
  {Vecchiato}, {Veljanoski}, {Via}, {Vicente}, {Vogt}, {Voss}, {Votruba},
  {Voutsinas}, {Walmsley}, {Weiler}, {Weingrill}, {Werner}, {Wevers},
  {Whitehead}, {Wyrzykowski}, {Yoldas}, {{\v{Z}}erjal}, {Zucker}, {Zurbach},
  {Zwitter}, {Alecu}, {Allen}, {Allende Prieto}, {Amorim},
  {Anglada-Escud{\'e}}, {Arsenijevic}, {Azaz}, {Balm}, {Beck}, {Bernstein},
  {Bigot}, {Bijaoui}, {Blasco}, {Bonfigli}, {Bono}, {Boudreault}, {Bressan},
  {Brown}, {Brunet}, {Bunclark}, {Buonanno}, {Butkevich}, {Carret}, {Carrion},
  {Chemin}, {Ch{\'e}reau}, {Corcione}, {Darmigny}, {de Boer}, {de Teodoro}, {de
  Zeeuw}, {Delle Luche}, {Domingues}, {Dubath}, {Fodor}, {Fr{\'e}zouls},
  {Fries}, {Fustes}, {Fyfe}, {Gallardo}, {Gallegos}, {Gardiol}, {Gebran},
  {Gomboc}, {G{\'o}mez}, {Grux}, {Gueguen}, {Heyrovsky}, {Hoar}, {Iannicola},
  {Isasi Parache}, {Janotto}, {Joliet}, {Jonckheere}, {Keil}, {Kim},
  {Klagyivik}, {Klar}, {Knude}, {Kochukhov}, {Kolka}, {Kos}, {Kutka}, {Lainey},
  {LeBouquin}, {Liu}, {Loreggia}, {Makarov}, {Marseille}, {Martayan},
  {Martinez-Rubi}, {Massart}, {Meynadier}, {Mignot}, {Munari}, {Nguyen},
  {Nordlander}, {Ocvirk}, {O'Flaherty}, {Olias Sanz}, {Ortiz}, {Osorio},
  {Oszkiewicz}, {Ouzounis}, {Palmer}, {Park}, {Pasquato}, {Peltzer}, {Peralta},
  {P{\'e}turaud}, {Pieniluoma}, {Pigozzi}, {Poels}, {Prat}, {Prod'homme},
  {Raison}, {Rebordao}, {Risquez}, {Rocca-Volmerange}, {Rosen}, {Ruiz-Fuertes},
  {Russo}, {Sembay}, {Serraller Vizcaino}, {Short}, {Siebert}, {Silva},
  {Sinachopoulos}, {Slezak}, {Soffel}, {Sosnowska}, {Strai{\v{z}}ys}, {ter
  Linden}, {Terrell}, {Theil}, {Tiede}, {Troisi}, {Tsalmantza}, {Tur},
  {Vaccari}, {Vachier}, {Valles}, {Van Hamme}, {Veltz}, {Virtanen}, {Wallut},
  {Wichmann}, {Wilkinson}, {Ziaeepour}, \& {Zschocke}}]{Prusti2016}
{Gaia Collaboration}, {Prusti}, T., {de Bruijne}, J.~H.~J., {et~al.} 2016,
  \aap, 595, A1, \dodoi{10.1051/0004-6361/201629272}

\bibitem[{{Gaia Collaboration} {et~al.}(2021){Gaia Collaboration}, {Brown},
  {Vallenari}, {Prusti}, {de Bruijne}, {Babusiaux}, {Biermann}, {Creevey},
  {Evans}, {Eyer}, {Hutton}, {Jansen}, {Jordi}, {Klioner}, {Lammers},
  {Lindegren}, {Luri}, {Mignard}, {Panem}, {Pourbaix}, {Randich}, {Sartoretti},
  {Soubiran}, {Walton}, {Arenou}, {Bailer-Jones}, {Bastian}, {Cropper},
  {Drimmel}, {Katz}, {Lattanzi}, {van Leeuwen}, {Bakker}, {Cacciari},
  {Casta{\~n}eda}, {De Angeli}, {Ducourant}, {Fabricius}, {Fouesneau},
  {Fr{\'e}mat}, {Guerra}, {Guerrier}, {Guiraud}, {Jean-Antoine Piccolo},
  {Masana}, {Messineo}, {Mowlavi}, {Nicolas}, {Nienartowicz}, {Pailler},
  {Panuzzo}, {Riclet}, {Roux}, {Seabroke}, {Sordo}, {Tanga}, {Th{\'e}venin},
  {Gracia-Abril}, {Portell}, {Teyssier}, {Altmann}, {Andrae}, {Bellas-Velidis},
  {Benson}, {Berthier}, {Blomme}, {Brugaletta}, {Burgess}, {Busso}, {Carry},
  {Cellino}, {Cheek}, {Clementini}, {Damerdji}, {Davidson}, {Delchambre},
  {Dell'Oro}, {Fern{\'a}ndez-Hern{\'a}ndez}, {Galluccio}, {Garc{\'\i}a-Lario},
  {Garcia-Reinaldos}, {Gonz{\'a}lez-N{\'u}{\~n}ez}, {Gosset}, {Haigron},
  {Halbwachs}, {Hambly}, {Harrison}, {Hatzidimitriou}, {Heiter},
  {Hern{\'a}ndez}, {Hestroffer}, {Hodgkin}, {Holl}, {Jan{\ss}en}, {Jevardat de
  Fombelle}, {Jordan}, {Krone-Martins}, {Lanzafame}, {L{\"o}ffler}, {Lorca},
  {Manteiga}, {Marchal}, {Marrese}, {Moitinho}, {Mora}, {Muinonen}, {Osborne},
  {Pancino}, {Pauwels}, {Petit}, {Recio-Blanco}, {Richards}, {Riello},
  {Rimoldini}, {Robin}, {Roegiers}, {Rybizki}, {Sarro}, {Siopis}, {Smith},
  {Sozzetti}, {Ulla}, {Utrilla}, {van Leeuwen}, {van Reeven}, {Abbas}, {Abreu
  Aramburu}, {Accart}, {Aerts}, {Aguado}, {Ajaj}, {Altavilla}, {{\'A}lvarez},
  {{\'A}lvarez Cid-Fuentes}, {Alves}, {Anderson}, {Anglada Varela}, {Antoja},
  {Audard}, {Baines}, {Baker}, {Balaguer-N{\'u}{\~n}ez}, {Balbinot}, {Balog},
  {Barache}, {Barbato}, {Barros}, {Barstow}, {Bartolom{\'e}}, {Bassilana},
  {Bauchet}, {Baudesson-Stella}, {Becciani}, {Bellazzini}, {Bernet}, {Bertone},
  {Bianchi}, {Blanco-Cuaresma}, {Boch}, {Bombrun}, {Bossini}, {Bouquillon},
  {Bragaglia}, {Bramante}, {Breedt}, {Bressan}, {Brouillet}, {Bucciarelli},
  {Burlacu}, {Busonero}, {Butkevich}, {Buzzi}, {Caffau}, {Cancelliere},
  {C{\'a}novas}, {Cantat-Gaudin}, {Carballo}, {Carlucci}, {Carnerero},
  {Carrasco}, {Casamiquela}, {Castellani}, {Castro-Ginard}, {Castro Sampol},
  {Chaoul}, {Charlot}, {Chemin}, {Chiavassa}, {Cioni}, {Comoretto}, {Cooper},
  {Cornez}, {Cowell}, {Crifo}, {Crosta}, {Crowley}, {Dafonte}, {Dapergolas},
  {David}, {David}, {de Laverny}, {De Luise}, {De March}, {De Ridder}, {de
  Souza}, {de Teodoro}, {de Torres}, {del Peloso}, {del Pozo}, {Delbo},
  {Delgado}, {Delgado}, {Delisle}, {Di Matteo}, {Diakite}, {Diener},
  {Distefano}, {Dolding}, {Eappachen}, {Edvardsson}, {Enke}, {Esquej}, {Fabre},
  {Fabrizio}, {Faigler}, {Fedorets}, {Fernique}, {Fienga}, {Figueras},
  {Fouron}, {Fragkoudi}, {Fraile}, {Franke}, {Gai}, {Garabato},
  {Garcia-Gutierrez}, {Garc{\'\i}a-Torres}, {Garofalo}, {Gavras}, {Gerlach},
  {Geyer}, {Giacobbe}, {Gilmore}, {Girona}, {Giuffrida}, {Gomel}, {Gomez},
  {Gonzalez-Santamaria}, {Gonz{\'a}lez-Vidal}, {Granvik},
  {Guti{\'e}rrez-S{\'a}nchez}, {Guy}, {Hauser}, {Haywood}, {Helmi}, {Hidalgo},
  {Hilger}, {H{\l}adczuk}, {Hobbs}, {Holland}, {Huckle}, {Jasniewicz},
  {Jonker}, {Juaristi Campillo}, {Julbe}, {Karbevska}, {Kervella}, {Khanna},
  {Kochoska}, {Kontizas}, {Kordopatis}, {Korn}, {Kostrzewa-Rutkowska},
  {Kruszy{\'n}ska}, {Lambert}, {Lanza}, {Lasne}, {Le Campion}, {Le Fustec},
  {Lebreton}, {Lebzelter}, {Leccia}, {Leclerc}, {Lecoeur-Taibi}, {Liao},
  {Licata}, {Lindstr{\o}m}, {Lister}, {Livanou}, {Lobel}, {Madrero Pardo},
  {Managau}, {Mann}, {Marchant}, {Marconi}, {Marcos Santos}, {Marinoni},
  {Marocco}, {Marshall}, {Martin Polo}, {Mart{\'\i}n-Fleitas}, {Masip},
  {Massari}, {Mastrobuono-Battisti}, {Mazeh}, {McMillan}, {Messina},
  {Michalik}, {Millar}, {Mints}, {Molina}, {Molinaro}, {Moln{\'a}r},
  {Montegriffo}, {Mor}, {Morbidelli}, {Morel}, {Morris}, {Mulone}, {Munoz},
  {Muraveva}, {Murphy}, {Musella}, {Noval}, {Ord{\'e}novic}, {Orr{\`u}},
  {Osinde}, {Pagani}, {Pagano}, {Palaversa}, {Palicio}, {Panahi}, {Pawlak},
  {Pe{\~n}alosa Esteller}, {Penttil{\"a}}, {Piersimoni}, {Pineau}, {Plachy},
  {Plum}, {Poggio}, {Poretti}, {Poujoulet}, {Pr{\v{s}}a}, {Pulone}, {Racero},
  {Ragaini}, {Rainer}, {Raiteri}, {Rambaux}, {Ramos}, {Ramos-Lerate}, {Re
  Fiorentin}, {Regibo}, {Reyl{\'e}}, {Ripepi}, {Riva}, {Rixon}, {Robichon},
  {Robin}, {Roelens}, {Rohrbasser}, {Romero-G{\'o}mez}, {Rowell}, {Royer},
  {Rybicki}, {Sadowski}, {Sagrist{\`a} Sell{\'e}s}, {Sahlmann}, {Salgado},
  {Salguero}, {Samaras}, {Sanchez Gimenez}, {Sanna}, {Santove{\~n}a},
  {Sarasso}, {Schultheis}, {Sciacca}, {Segol}, {Segovia}, {S{\'e}gransan},
  {Semeux}, {Shahaf}, {Siddiqui}, {Siebert}, {Siltala}, {Slezak}, {Smart},
  {Solano}, {Solitro}, {Souami}, {Souchay}, {Spagna}, {Spoto}, {Steele},
  {Steidelm{\"u}ller}, {Stephenson}, {S{\"u}veges}, {Szabados}, {Szegedi-Elek},
  {Taris}, {Tauran}, {Taylor}, {Teixeira}, {Thuillot}, {Tonello}, {Torra},
  {Torra}, {Turon}, {Unger}, {Vaillant}, {van Dillen}, {Vanel}, {Vecchiato},
  {Viala}, {Vicente}, {Voutsinas}, {Weiler}, {Wevers}, {Wyrzykowski}, {Yoldas},
  {Yvard}, {Zhao}, {Zorec}, {Zucker}, {Zurbach}, \& {Zwitter}}]{GaiaEDR3}
{Gaia Collaboration}, {Brown}, A.~G.~A., {Vallenari}, A., {et~al.} 2021, \aap,
  649, A1, \dodoi{10.1051/0004-6361/202039657}

\bibitem[{{Lindegren} {et~al.}(2018){Lindegren}, {Hern{\'a}ndez}, {Bombrun},
  {Klioner}, {Bastian}, {Ramos-Lerate}, {de Torres}, {Steidelm{\"u}ller},
  {Stephenson}, {Hobbs}, {Lammers}, {Biermann}, {Geyer}, {Hilger}, {Michalik},
  {Stampa}, {McMillan}, {Casta{\~n}eda}, {Clotet}, {Comoretto}, {Davidson},
  {Fabricius}, {Gracia}, {Hambly}, {Hutton}, {Mora}, {Portell}, {van Leeuwen},
  {Abbas}, {Abreu}, {Altmann}, {Andrei}, {Anglada}, {Balaguer-N{\'u}{\~n}ez},
  {Barache}, {Becciani}, {Bertone}, {Bianchi}, {Bouquillon}, {Bourda},
  {Br{\"u}semeister}, {Bucciarelli}, {Busonero}, {Buzzi}, {Cancelliere},
  {Carlucci}, {Charlot}, {Cheek}, {Crosta}, {Crowley}, {de Bruijne}, {de
  Felice}, {Drimmel}, {Esquej}, {Fienga}, {Fraile}, {Gai}, {Garralda},
  {Gonz{\'a}lez-Vidal}, {Guerra}, {Hauser}, {Hofmann}, {Holl}, {Jordan},
  {Lattanzi}, {Lenhardt}, {Liao}, {Licata}, {Lister}, {L{\"o}ffler},
  {Marchant}, {Martin-Fleitas}, {Messineo}, {Mignard}, {Morbidelli}, {Poggio},
  {Riva}, {Rowell}, {Salguero}, {Sarasso}, {Sciacca}, {Siddiqui}, {Smart},
  {Spagna}, {Steele}, {Taris}, {Torra}, {van Elteren}, {van Reeven}, \&
  {Vecchiato}}]{Lindegren2018}
{Lindegren}, L., {Hern{\'a}ndez}, J., {Bombrun}, A., {et~al.} 2018, \aap, 616,
  A2, \dodoi{10.1051/0004-6361/201832727}

\bibitem[{{Lubow} {et~al.}(2021){Lubow}, {White}, \& {Shiao}}]{Lubow2021}
{Lubow}, S.~H., {White}, R.~L., \& {Shiao}, B. 2021, \aj, 161, 6,
  \dodoi{10.3847/1538-3881/abc267}

\bibitem[{{Magnier} {et~al.}(2020){Magnier}, {Schlafly}, {Finkbeiner}, {Tonry},
  {Goldman}, {R{\"o}ser}, {Schilbach}, {Casertano}, {Chambers}, {Flewelling},
  {Huber}, {Price}, {Sweeney}, {Waters}, {Denneau}, {Draper}, {Hodapp},
  {Jedicke}, {Kaiser}, {Kudritzki}, {Metcalfe}, {Stubbs}, \&
  {Wainscoat}}]{Magnier2020}
{Magnier}, E.~A., {Schlafly}, E.~F., {Finkbeiner}, D.~P., {et~al.} 2020, \apjs,
  251, 6, \dodoi{10.3847/1538-4365/abb82a}

\bibitem[{{Tachibana} \& {Miller}(2018)}]{Tachibana2018}
{Tachibana}, Y., \& {Miller}, A.~A. 2018, \pasp, 130, 128001,
  \dodoi{10.1088/1538-3873/aae3d9}

\end{thebibliography}



\end{document}